\documentclass[manuscript]{emulateapj}

\usepackage{amssymb}
\usepackage{amsmath}
\usepackage{booktabs}
\usepackage{hyperref}
\usepackage{multirow}
\usepackage{times}
\PassOptionsToPackage{hyphens}{url}\usepackage{hyperref}
\usepackage[hyphenbreaks]{breakurl}

\newcommand{\ls}
{\mathrel{\hbox{\rlap{\hbox{\lower4pt\hbox{$\sim$}}}\hbox{$<$}}}}
\newcommand{\gs}
{\mathrel{\hbox{\rlap{\hbox{\lower4pt\hbox{$\sim$}}}\hbox{$>$}}}}

\received{}
\begin{document}
\title{The variable, fast soft X-ray wind in PG\,1211+143}
\shorttitle{The soft X-ray wind in PG\,1211+143}
\shortauthors{Reeves et al.}
\author{J. N. Reeves\altaffilmark{1}, A. Lobban\altaffilmark{2}, K. A. Pounds\altaffilmark{3}} 

\altaffiltext{1}{Center for Space Science and Technology, 
University of Maryland Baltimore County, 1000 Hilltop Circle, Baltimore, MD 21250, USA; email jreeves@umbc.edu}
\altaffiltext{2}{Astrophysics Group, School of Physical and Geographical Sciences, Keele 
University, Keele, Staffordshire, ST5 5BG, UK}
\altaffiltext{3}{Dept of Physics and Astronomy, University of Leicester, University Road, Leicester LE1 7RH, UK}

\begin{abstract}

The analysis of a series of seven observations of the nearby ($z=0.0809$) QSO, PG\,1211+143, taken with the Reflection Grating Spectrometer (RGS) on-board {\it XMM-Newton} in 2014, are presented. The high resolution soft X-ray spectrum, with a total exposure exceeding 600\,ks, shows a series of blue-shifted absorption lines, from the He and H-like transitions of N, O and Ne, as well as from L-shell Fe. 
The strongest absorption lines are all systematically blue-shifted by $-0.06c$, 
originating in two absorption zones, from low and high ionization gas. 
Both zones are variable on timescales of days, with the variations in absorber opacity effectively explained by either column density changes  
or by the absorber ionization responding directly to the continuum flux. We find that the soft X-ray absorbers probably exist in a two-phase wind, at a radial distance of $\sim10^{17}-10^{18}$\,cm from the black hole, 
with the lower ionization gas as denser clumps embedded within a higher ionization outflow. 
The overall mass outflow rate of the soft X-ray wind may be as high as $2{\rm M}_{\odot}$\,yr$^{-1}$, 
close to the Eddington rate for PG\,1211+143 and similar to that previously deduced from the Fe K 
absorption.

\end{abstract}










\section{Introduction}

The X-ray spectra of active galactic nuclei (AGN) are commonly observed to display evidence for significant columns of ionized circumnuclear material within the vicinity of the central black hole.  In particular, high-resolution X-ray spectra acquired with {\it XMM-Newton} and {\it Chandra} have revealed the presence of narrow, blueshifted absorption lines, indicative of outflowing material in our line of sight (e.g. \citealt{Kaastra00, Kaspi02, CrenshawKraemerGeorge03, McKernanYaqoobReynolds07}).  With the absorbing material typically outflowing with velocities on the order of a few times $-100$\,km\,s$^{-1}$ to  $\sim -1\,000$\,km\,s$^{-1}$, these so-called ``warm absorbers'' are observed to be present in at least 50\,per cent of AGN (e.g. \citealt{ReynoldsFabian95, Blustin05}).

In addition to the warm absorbers, typically observed in the soft X-ray band, the discovery of blueshifted absorption features from much more highly-ionized material has provided strong evidence for the existence of high-velocity outflows.  With the signature primarily manifesting itself in absorption lines from K-shell transitions of He-/H-like Fe, the measured blueshifts imply outflow velocities on the order of $\sim-0.1c$ or higher (e.g. APM\,08279+5255; \citealt{Chartas02}, PG\,1211+143; \citealt{Pounds03}, PDS\,456; \citealt{ReevesOBrienWard03}).  The high velocities and ionization of the material suggests that the high-velocity outflows originate much closer to the black hole than the slower, less-ionized warm absorbers.  More recent systematic studies of archival {\it XMM-Newton} \citep{Tombesi10, Tombesi11} and {\it Suzaku} \citep{Gofford13} data have shown that high-velocity disc winds may be a common feature of luminous AGN.  With the derived outflow rates calculated to be high (up to $\sim$few $M_{\odot}$\,yr$^{-1}$), and therefore comparable with the measured accretion rates of AGN, the outflows are estimated to carry kinetic power as much as a few per cent of the bolometric luminosity, $L_{\rm bol}$.  As such, fast outflows likely play an important role in coupling together black-hole growth and the properties of the host galaxy \citep{King03, King10}, and offer a possible interpretation of the $M-\sigma$ relation for galaxies \citep{FerrareseMerritt00, Gebhardt00}.

While the evidence for high-velocity outflows in the hard X-ray / Fe\,K band is now well established, there is a relative scarcity of detections to date in the soft X-ray band.  Detecting lower-ionization, soft-band counterparts of high-velocity outflows is important, not only to fully establish the presence of high-velocity outflows in AGN, but also to probe all phases of the outflowing gas -- i.e. across a wide range of ionization and column density.

In addition to the original discovery of the fast wind in PG\,1211+143 \citep{Pounds03}, evidence has emerged for the presence of a fast soft X-ray absorber in several other AGN.
Notably in PDS\,456, \citet{ReevesOBrienWard03} first noted the presence of blueshifted absorption in the soft X-ray band through an {\it XMM-Newton} observation in 2001. Here a broad absorption trough was resolved in the RGS spectrum, originating from a blend of L-shell transitions from highly-ionized Fe near $\sim$1\,keV, with a measured outflow velocity of $\sim -50\,000$\,km\,s$^{-1}$ (in addition to the now well-established high-velocity outflow detected in the Fe\,K band).  These high-velocity, soft-band counterparts of the outflow have since been confirmed in a multi-epoch analysis of all of the {\it XMM-Newton} observations of PDS\,456 \citep{Reeves16}.

Further claims of detections of high-velocity outflows in the soft X-ray band include Ark\,564 \citep{Gupta13}, where K$\alpha$ transitions of O\,\textsc{vii} and O\,\textsc{vi} were detected in absorption with a measured blueshift corresponding to $v_{\rm out} \sim -0.1c$, and Mrk\,590 \citep{GuptaMathurKrongold15}, primarily through the detection of blueshifted absorption lines from O\,\textsc{viii}, Ne\,\textsc{ix}, Si\,\textsc{xiv} and Mg\,\textsc{xii}, ranging in outflow velocity from $-0.07c$ to $-0.18c$.  Both observations were performed with the gratings on-board {\it Chandra}.

More recently, multiple-velocity components of a fast outflow were observed in the {\it XMM-Newton} RGS spectrum of the narrow-line Seyfert 1 galaxy IRAS\,17020+4544 \citep{Longinotti15}, covering a wide range of ionization and column density with outflow velocities in the range $-23\,000$ to $-33\,000$\,km\,s$^{-1}$.  A soft-band absorber with a velocity of $\sim -0.24c$ was also observed in IRAS\,13224-3809 \citep{Parker17} through high-resolution RGS data with a corresponding highly-ionized Fe\,K counterpart and a claim of a correlation between the observed properties of the outflow and the luminosity of the source \citep{Pinto17a}.  A wealth of absorption lines were also observed in the RGS spectrum of NGC\,4051 with velocity components of up to $\sim -10\,000$\,km\,s$^{-1}$ \citep{PoundsVaughan11}, interpreted as a cooling shocked flow 
\citep{PoundsKing13}.
Finally, in addition to such detections in AGN, evidence for high-velocity outflows in the soft X-ray band has also been seen in ultra-luminous X-ray sources (ULXs) -- e.g. in NGC\,1313 X-1 and NGC\,5408 X-1 \citep{Pinto17b, Pinto17c}, with measured outflow velocities of $\sim -0.2c$.

This paper focuses on PG\,1211+143, which is a luminous narrow-line Seyfert galaxy / quasi-stellar object (QSO), at a distance of 331\,Mpc ($z = 0.0809$; \citealt{Marziani96}).  It is both bright in the optical band and the X-ray band with an X-ray luminosity of the order of $\sim$10$^{44}$\,erg\,s$^{-1}$.  While being known for its variability and spectral complexity, PG\,1211+143 is primarily known for being the `prototype' source for high-velocity outflows as it offered the first detection of a mildly-relativistic outflow in a non-broad-absorption-line (BAL) AGN \citep{Pounds03, PoundsPage06}, with a velocity of $v_{\rm out} \sim -0.1c$.

The initial detection of the high-velocity outflow was based on the detection of blueshifted absorption in the hard X-ray / Fe\,K band, from the original 60\,ks {\it XMM-Newton} observations in 2001. 
In addition to the Fe K band absorption, \citet{Pounds03} also claimed the detection of several 
blueshifted absorption lines in the RGS spectra, originating from the He and H-like lines of C, N, O and Ne.  
Although \citet{KaspiBehar06} claimed a much lower velocity, 
a more recent analysis combining multiple resonance lines confirmed the soft X-ray outflow velocity of 
$-0.07c$ \citep{Pounds14}. 
Photoionized absorption and emission spectra from 3 XMM-Newton observations were then modeled over a broad spectral band, to estimate the mass outflow rate and energetics of the flow, suggesting that the outflow was significant in terms of galactic feedback \citep{PoundsReeves07, PoundsReeves09, Pounds14b}.

\begin{deluxetable*}{lcccccc}
\tablecaption{Observation log of the seven {\it XMM-Newton} RGS observations of PG\,1211+143 in 2014.  Count rates and mean fluxes are given over the 0.4--2\,keV band.}
\tablewidth{0pt}
\tablehead{
\colhead{Start Date} & \colhead{Obs. ID} & \colhead{{\it XMM} Rev.} & \colhead{Instrument} & \colhead{Exposure$^{a}$} & \colhead{Count Rate$^{b}$} & \colhead{Mean Flux$^{c}$} \\
}
\startdata
\multirow{2}{*}{2014-06-02} & \multirow{2}{*}{0745110101} & \multirow{2}{*}{\textsc{rev}\,2652} & RGS\,1 & 85.7 & $0.137 \pm 0.001$ & $5.41^{+0.10}_{-0.11}$  \\
&  &  & RGS\,2 & 85.6 & $0.136 \pm 0.001$ & $5.36^{+0.12}_{-0.10}$  \\
\multirow{2}{*}{2014-06-15} & \multirow{2}{*}{0745110201} & \multirow{2}{*}{\textsc{rev}\,2659} & RGS\,1 & 102.7 & $0.086 \pm 0.001$ & $3.46^{+0.07}_{-0.12}$ \\
& & & RGS\,2 & 102.8 & $0.088 \pm 0.001$ & $3.40^{+0.08}_{-0.08}$ \\
\multirow{2}{*}{2014-06-19} & \multirow{2}{*}{0745110301} & \multirow{2}{*}{\textsc{rev}\,2661} & RGS\,1 & 101.1 & $0.130 \pm 0.001$ & $5.00^{+0.11}_{-0.09}$ \\
& & & RGS\,2 & 101.2 & $0.126 \pm 0.001$ & $4.68^{+0.07}_{-0.09}$ \\
\multirow{2}{*}{2014-06-23} & \multirow{2}{*}{0745110401} & \multirow{2}{*}{\textsc{rev}\,2663} & RGS\,1 & 98.7 & $0.136 \pm 0.001$ & $5.29^{+0.09}_{-0.09}$ \\
& & & RGS\,2 & 98.8 & $0.133 \pm 0.001$ & $4.84^{+0.07}_{-0.08}$ \\
\multirow{2}{*}{2014-06-25} & \multirow{2}{*}{0745110501} & \multirow{2}{*}{\textsc{rev}\,2664} & RGS\,1 & 56.8 & $0.181 \pm 0.002$ & $6.90^{+0.14}_{-0.13}$ \\
& & & RGS\,2 & 56.8 & $0.182 \pm 0.001$ & $6.60^{+0.12}_{-0.12}$ \\
\multirow{2}{*}{2014-06-29} & \multirow{2}{*}{0745110601} & \multirow{2}{*}{\textsc{rev}\,2666} & RGS\,1 & 94.0 & $0.169 \pm 0.001$ & $6.46^{+0.14}_{-0.06}$ \\
& & & RGS\,2 & 94.0 & $0.173 \pm 0.001$ &$6.27^{+0.10}_{-0.13}$ \\
\multirow{2}{*}{2016-07-07} & \multirow{2}{*}{0745110701} & \multirow{2}{*}{\textsc{rev}\,2670} & RGS\,1 & 97.6 & $0.134 \pm 0.001$ & $5.24^{+0.11}_{-0.06}$ \\
& & & RGS\,2 & 97.7 & $0.132 \pm 0.001$ & $4.85^{+0.07}_{-0.09}$ \\
\midrule
-- & -- & mean & RGS\,1+2 & 1274 & $0.136\pm0.001$ & $5.07^{+0.01}_{-0.01}$ \\
\enddata
\tablenotetext{a}{Net exposure time in units of ks.}
\tablenotetext{b}{Observed background-subtracted count rate in units of ct\,s$^{-1}$.}
\tablenotetext{c}{Observed mean flux in units of $\times 10^{-12}$\,erg\,cm$^{-2}$\,s$^{-1}$.}
\label{tab:rgs_obs_log}
\end{deluxetable*}

This paper marks the sixth in a series of papers on an extended $\sim$630\,ks {\it XMM-Newton} observation acquired in 2014.  In \citet{Pounds16a}, an analysis of the hard X-ray spectrum revealed significant velocity structure in the highly-ionized wind, with primary velocities of $v/c \sim -0.06$ and $v/c \sim -0.13c$.  In \citet{Pounds16b}, we extended the analysis to the soft X-ray band through the high signal-to-noise Reflection Grating Spectrometer (RGS) spectrum, detecting lower-ionization high-velocity counterparts of the outflow.  Meanwhile, the multi-wavelength variability with {\it XMM-Newton} and a long-term {\it Swift} monitoring campaign was analyzed in \citet{Lobban16a}, while the broad-band {\it XMM-Newton} $+$ {\it NuSTAR} spectrum was analyzed in \citet{Lobban16b}, showing that the high-velocity wind model extends well to the broad 0.3--79\,keV bandpass.  Finally, in \citet{Lobban17} we analyze the short-term X-ray variability of the source through inter-band Fourier time lags.  

In this paper, we revisit the mean RGS spectrum from the 
{\it XMM-Newton} long look, finding the dominant soft X-ray absorption arises from low and high ionization gas, outflowing with a common velocity of $-0.06c$.
We then focus on the short-term variability of the outflow in the soft X-ray band on the timescale of $\sim$days through the analysis of high signal-to-noise inter-orbit RGS spectra, finding substantial short-term variability in the ultra fast outflow in PG\,1211+143.

\section{Observations and Data Reduction}

PG\,1211+143 was observed by {\it XMM-Newton} \citep{Jansen01} on seven occasions in 2014.  The observations took place over a roughly five-week period from 2014-06-02 to 2014-07-07.  The duration of the seven observations totalled $\sim$630\,ks with each observation having a typical duration of $\sim$100\,ks, apart from the fifth observation which was $\sim$55\,ks.  In this paper, we primarily focus on data acquired with the RGS \citep{denHerder01} on-board {\it XMM-Newton}.  All data were processed using the {\it XMM-Newton} Scientific Analysis Software (\textsc{sas}\footnote{\url{http://xmm.esac.esa.int/sas/}}) package (version 15.0).

The \textsc{rgsproc}\footnote{\url{http://xmm-tools.cosmos.esa.int/external/sas/current/doc/rgsproc/}} script as part of \textsc{sas} was used to extract first-order dispersed spectra from each of the two RGS modules.  Observations were screened for periods of high background by examining lightcurves from the CCD closest to the optical axis of the telescope, 
resulting in net exposures ranging from 56.8\,ks (per RGS module) for the shortest observation to 102.8\,ks.  The net background-subtracted count rates range from $0.086 \pm 0.001$ to $0.181 \pm 0.002$\,ct\,s$^{-1}$ for RGS\,1 and $0.088 \pm 0.001$ to $0.182 \pm 0.001$\,ct\,s$^{-1}$ for RGS\,2.  These translate into 0.4--2\,keV fluxes of $3.40^{+0.07}_{-0.12} \times 10^{-12}$ to $6.90^{+0.14}_{-0.13} \times 10^{-12}$  and $3.40 \pm 0.08 \times 10^{-12}$ to $6.60 \pm 0.12 \times 10^{-12}$\,erg\,cm$^{-2}$\,s$^{-1}$ for the two detectors, respectively.  In all seven observations, the background count rate was just a few per cent of the source rate. The exception to this is the very long wavelength part of the RGS spectrum above 30\,\AA, as well as the very short wavelength band below $\sim7.5$\,\AA\ in the observed frame, where the effective area drops rapidly. Thus the subsequent spectral analysis is restricted to $7.5-30$\,\AA\ in the observed frame.  
All count rates and observed fluxes are listed in Table~\ref{tab:rgs_obs_log}.

\begin{figure*}
\begin{center}
\rotatebox{0}{\includegraphics[width=15cm]{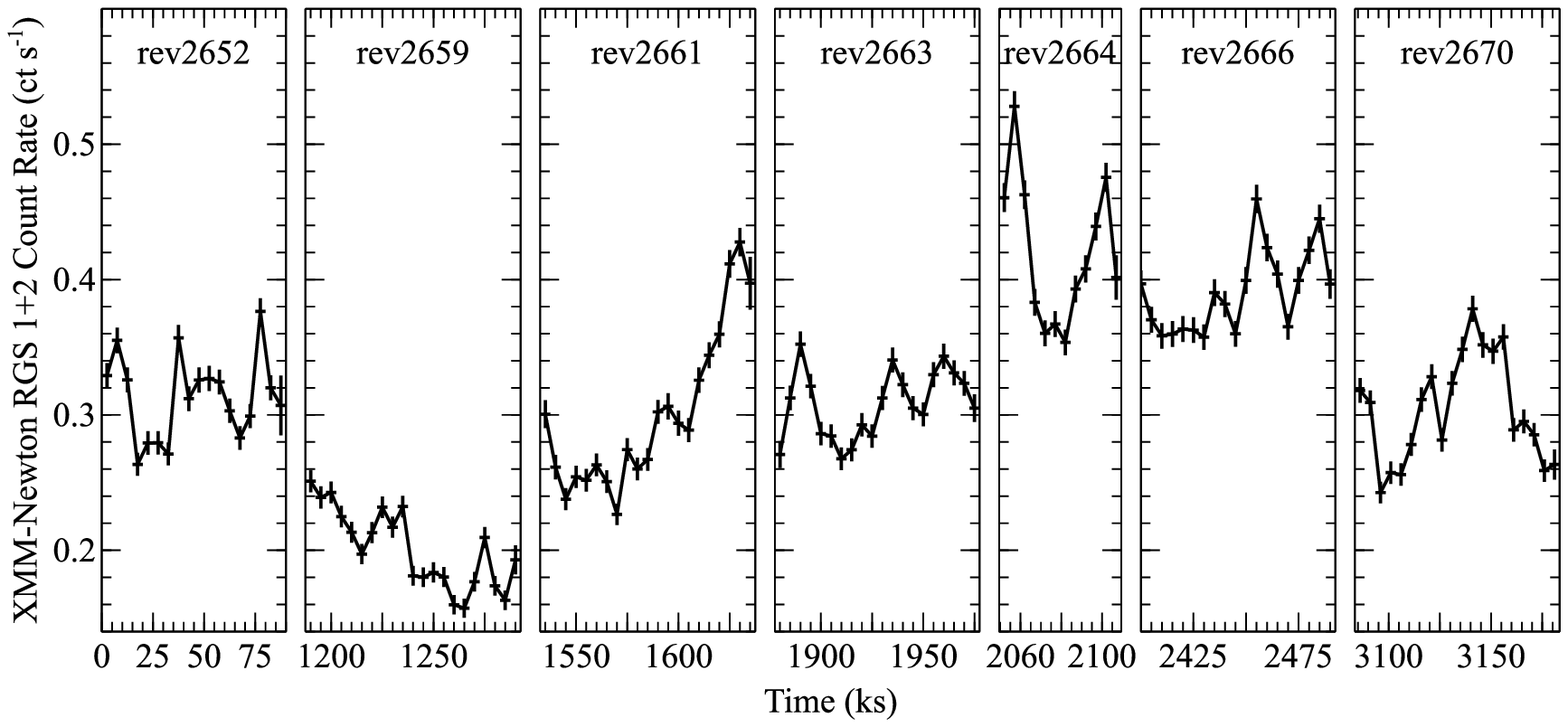}}
\end{center}
\caption{The {\it XMM-Newton} first-order 0.4--2\,keV RGS\,$1+2$ light curve of PG\,1211+143 in 5\,ks time bins. The lightcurve shows the relative variability between all seven {\it XMM-Newton} sequences in the soft band, with \textsc{rev}\,2659 the lowest-flux observation.
Note the gaps between the observations, where the x-axis corresponds to time since the beginning of the campaign in \textsc{rev}\,2652.  See Table~\ref{tab:rgs_obs_log} for the 
details of the individual {\it XMM-Newton} sequences.}
\label{fig:rgs_lightcurve}
\end{figure*}

The RGS\,1 and RGS\,2 spectra were found to be consistent in each of the seven observations and 
the subsequent spectral parameters were found to be consistent from fitting RGS\,1+2 separately. 
As such, combined RGS\,$1+2$ spectra were produced, using the \textsc{rgscombine}\footnote{\url{http://xmm-tools.cosmos.esa.int/external/sas/current/doc/rgscombine/}} task within \textsc{sas}.  Finally, we also combined the spectra from all seven observations to produce a high signal-to-noise time-averaged `mean' RGS\,$1+2$ spectrum, having weighted the response files by exposure time.  This spectrum has a mean count rate of $0.136 \pm 0.001$\,ct\,s$^{-1}$ and 0.4--2\,keV flux of $5.07 \pm 0.01 \times 10^{-12}$\,erg\,cm$^{-2}$\,s$^{-1}$.  All spectra were binned using the \textsc{sas} task \textsc{specgroup}\footnote{\url{https://xmm-tools.cosmos.esa.int/external/sas/current/doc/specgroup/}} in such a way as to not oversample the intrinsic energy resolution of the detector(s) by any factor greater than 3.  This also ensured that there were $> 25$\,ct\,bin$^{-1}$, thus allowing the use of $\chi^{2}$ minimization. Note that the combined spectrum yielded a total of 173,000 net counts, 
while for the individual sequences this ranges from 18,000 counts (rev\,2659) to 
32,000 counts (rev 2666), sufficient for the use of $\chi^{2}$ statistics.
Errors are quoted at 90\,per cent confidence for one parameter of interest ($\Delta \chi^{2} = 2.7$).  All subsequent fits are performed using the \textsc{xspec} spectral fitting package \citep{Arnaud96}.

\begin{figure}
\begin{center}
\rotatebox{0}{\includegraphics[width=8.5cm]{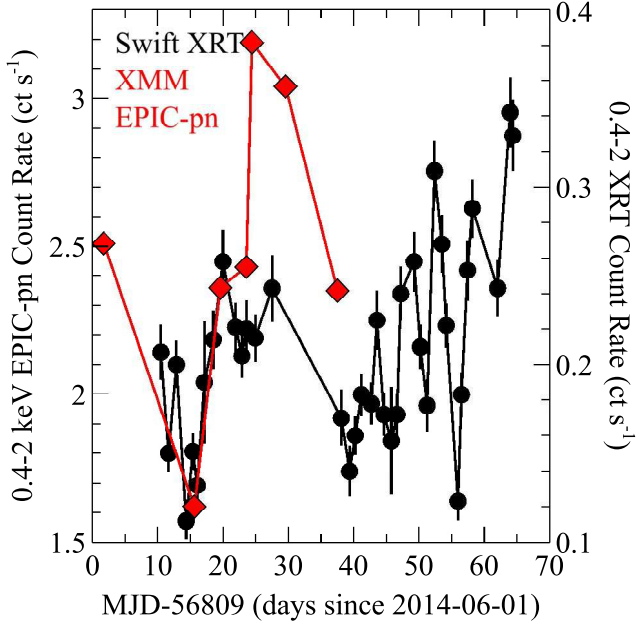}}
\end{center}
\caption{The overlaid {\it XMM-Newton} EPIC-pn (red diamonds) and {\it Swift} XRT (black circles) lightcurve in the 0.4--2\,keV band, showing the long-term soft X-ray variability of PG\,1211+143. Note the second {\it XMM-Newton} point in \textsc{rev}\,2659 occurs during a period of low soft X-ray flux $\sim$15\,days 
after the start of the observational campaign. The soft X-ray flux then increases, peaking in {\it XMM-Newton} \textsc{rev}\,2664.}
\label{fig:xrt_pn_lc_flux}
\end{figure}

\section{Mean Spectral Analysis}

In Figure~\ref{fig:rgs_lightcurve}, we show the combined RGS\,$1+2$ lightcurve of PG\,1211+143 over the soft 0.4--2\,keV band.  The source is observed to be highly variable in the soft band, varying in count rate by up to a factor of $\sim$3 from as low as $\sim$0.15\,ct\,s$^{-1}$ during \textsc{rev}\,2659 to $\sim$0.5\,ct\,s$^{-1}$ in \textsc{rev}\,2664.  In Figure~\ref{fig:xrt_pn_lc_flux}, the soft-band lightcurve is placed into the context of a longer $\sim$2-month monitoring campaign which we obtained with {\it Swift} \citep{Gehrels04}, where we have overlaid the observed 0.4--2\,keV {\it XMM-Newton} EPIC-pn and {\it Swift} XRT count rates.  The broad-band {\it XMM-Newton} and {\it Swift} variability is described in \citet{Lobban16a} and we return to the soft-band spectral variability of PG\,1211+143 later in Section~\ref{sec:outflow_variability}.  Our initial aim is to build a template model based on the mean spectrum of PG\,1211+143 from the 2014 RGS observations, which then can be applied to all seven sequences (listed in Table~\ref{tab:rgs_obs_log}) to study the variability of the fast soft X-ray wind. We start by revisiting the spectral features in the mean spectrum and their common outflow velocity, subsequently building upon that to construct a best-fitting photoionization model. 

Note that outflow velocities are given in the rest-frame of PG\,1211+143 at $z=0.0809$, 
accounting for relativistic Doppler shifts, where negative values of $v/c$ correspond to blue-shifted velocities. 
In order to calculate the outflow velocity ($v_{\rm out}$) of an absorption line system in the rest frame of PG\,1211+143, the following prescription is used:-
\begin{equation}
\frac{v_{\rm out}}{c} = \frac{(1 + z_{\rm obs})^2 - (1 + z_{\rm QSO})^2}{(1 + z_{\rm obs})^2 + (1 + z_{\rm QSO})^2}
\end{equation}
where $z_{\rm obs}$ is the redshift of an absorption system in the observed frame and 
$z_{\rm QSO}$ is the QSO redshift ($z=0.0809$).

\subsection{Soft X-ray Absorption and Emission lines from PG\,1211+143} \label{sec:soft_x-ray_absorption_emission}

To investigate the mean soft X-ray spectrum, which was initially analyzed in \citet{Pounds16b}, 
we constructed a simple baseline continuum model which we applied to the RGS spectrum 
over the rest frame range 7--27.5\,\AA\ (or 7.5--30\,\AA\ observed frame), which covers the highest S/N portion of the spectrum.
A two-component continuum model was adopted consisting of 
two power laws, one of which is steep ($\Gamma\sim3$) to account for the prominent soft excess in PG\,1211+143 and a harder component ($\Gamma=1.6-1.7$) 
which dominates the broad-band spectrum above 1\,keV \citep{Lobban16a}. 
A neutral photoelectric absorber, associated with our Galaxy, was also included via the \textsc{xspec} \textsc{tbabs} model \citep{WilmsAllenMcCray00}, using the absorption cross-sections of \citet{Verner96} 
and the abundances of \citet{WilmsAllenMcCray00}. This was initially set to the expected column density of $N_{\rm H}=2.9\times10^{20}$\,cm$^{-2}$ \citep{Kalberla05}, but was subsequently allowed to vary to allow for any column density in excess of this value.

\begin{figure*}
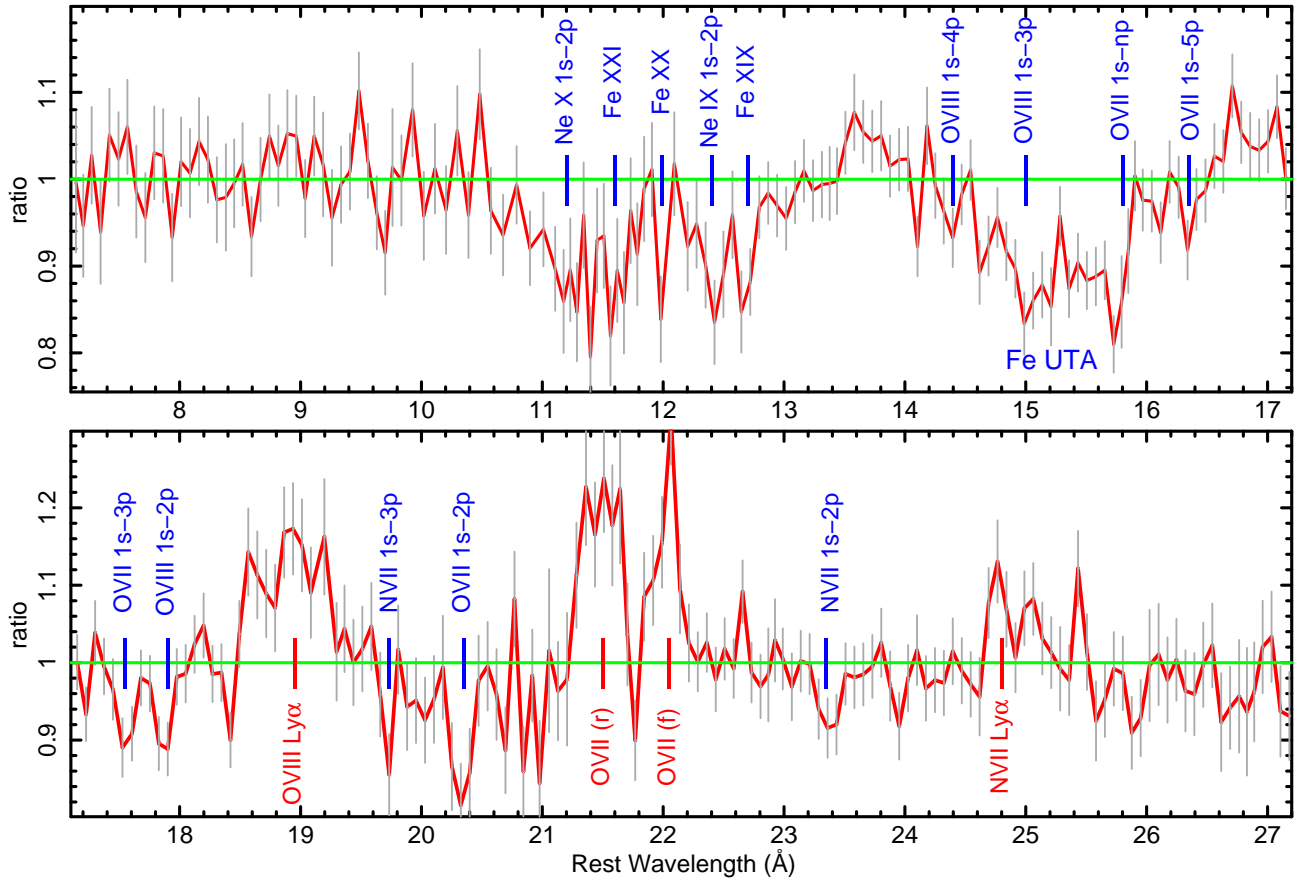

\begin{center}
\rotatebox{-90}{\includegraphics[height=18cm]{f3a.eps}}
\rotatebox{-90}{\includegraphics[height=18cm]{f3b.eps}}
\end{center}
\caption{The mean {\it XMM-Newton} RGS\,$1+2$ spectrum of PG\,1211+143, with a total exposure of $\sim640$\,ks, plotted as a ratio to the baseline double power law continuum model. The data are binned by a factor of 
$\times 2$ for clarity.
The upper panel shows the 7--17\,\AA\ band and the 
lower panel from 17--27\,\AA\ in the quasar rest frame at $z=0.0809$. 
Data are shown as red histograms, while $1\sigma$ error 
bars are in grey. Several strong resonance ($1s \rightarrow 2p$) 
absorption lines from Ne, O, N, as well as L-shell Fe, are present in the spectrum. These are systematically blueshifted with respect to their known wavelengths, by an outflow 
velocity of $-0.06$ up to $-0.08c$ (see Table~\ref{tab:absorption-lines}). The broad absorption trough at $\sim$15\,\AA\ is due to a blueshifted Fe M-shell Unresolved Transition Array (UTA). 
Note also the presence of strong and broadened emission lines, especially due to the O\,\textsc{vii} triplet (21.6--22.1\AA) and O\,\textsc{viii} Ly$\alpha$ (19\AA).}
\label{fig:mean_ratio}
\end{figure*}

The data/model residuals to this baseline continuum are shown in Figure~\ref{fig:mean_ratio}, where all the significant lines 
occur over the wavelength range from 7--27\,\AA\ in the quasar rest frame (or up to 30\,\AA\ in the 
observed frame). Several prominent 
narrow absorption lines are present in the soft X-ray spectrum, primarily arising from the resonance 
He and H-like lines of N, O and Ne, as well as from L-shell Fe. In Table~\ref{tab:absorption-lines} we list the centroid 
rest wavelengths of the most prominent absorption lines observed in the spectrum (detected at a minimum 95\,per cent confidence level), along with their most likely atomic identification and corresponding laboratory frame wavelength. 


In total, there are 14 narrow lines (plus one edge) 
in Table~\ref{tab:absorption-lines}, which can be identified with absorption at a consistent velocity, with a 
mean velocity shift of $-0.061\pm0.001c$. However, two lines 
may have uncertain identifications. 
The absorption line near $\sim$17.5\,\AA\ 
may either be associated to O\,\textsc{vii} $1s\rightarrow 3p$ at $-0.06c$, or a higher-velocity
component of O\,\textsc{viii} $1s\rightarrow 2p$ at $-0.08c$. In addition, the absorption line at 12.4\,\AA\ could also be associated with a higher ($-0.08c$) velocity component of Ne\,\textsc{ix} $1s\rightarrow 2p$, with no obvious lower velocity counterpart. 
This supports the earlier Gaussian analysis presented in \citet{Pounds16b}, confirming the presence of a soft X-ray counterpart to the wind which was originally claimed in \citet{Pounds03} on the basis of the 
initial 60\,ks {\it XMM-Newton} observation in 2001.

\begin{deluxetable}{lcccccc}
\tablecaption{Blueshifted absorption lines identified in the mean 2014 RGS spectrum of PG\,1211+143.}
\tablewidth{0pt}
\tablehead{
\colhead{Rest $\lambda$ (\AA)$^{a}$} & \colhead{Line ID$^{b}$} & \colhead{Lab $\lambda$ (\AA)$^{b}$} & \colhead{v/c$^{c}$}}
\startdata
23.35 & N\,\textsc{vii} $1s\rightarrow2p$ & 24.78 & $-0.059\pm0.002$ \\
20.33 & O\,\textsc{vii} $1s\rightarrow2p$ &  21.60 & $-0.060\pm0.002$ \\
19.71 & N\,\textsc{vii} $1s\rightarrow3p$ & 20.91 & $-0.059\pm0.002$ \\
17.86 & O\,\textsc{viii} $1s\rightarrow2p$ & 18.97 & $-0.060\pm0.002$ \\
17.54 & O\,\textsc{vii} $1s\rightarrow3p$ & 18.63 & $-0.060\pm0.002$ \\
$-$ & or O\,\textsc{viii} $1s\rightarrow2p$ & 18.97 &  $-0.078\pm0.002^*$ \\
16.32 & O\,\textsc{vii} $1s\rightarrow5p$ & 17.40 & $-0.064\pm0.002$ \\
16.10 & O\,\textsc{vii} $1s\rightarrow6p$ & 17.20 & $-0.066\pm0.002$ \\
15.80 & O\,\textsc{vii} edge & 16.77 & $-0.059\pm0.002$ \\
14.99 & O\,\textsc{viii} $1s\rightarrow3p$ & 16.00 & $-0.065\pm0.003$ \\
14.29 & O\,\textsc{viii} $1s\rightarrow4p$ & 15.18 & $-0.060\pm0.003$ \\
12.65 & Ne\,\textsc{ix} $1s\rightarrow2p$ & 13.45 & $-0.061\pm0.003$ \\
$-$ & or Fe\,\textsc{xix} $2p\rightarrow3d$ & 13.52 & $-0.064\pm0.004$ \\
12.41 & Ne\,\textsc{ix} $1s\rightarrow2p$ & 13.45 & $-0.080\pm0.003^*$ \\
11.99 & Fe\,\textsc{xx} $2p\rightarrow3d$ &12.82 & $-0.066\pm0.004$ \\
11.55 & Fe\,\textsc{xxi} $2p\rightarrow3d$ & 12.28 & $-0.061\pm0.004$ \\
11.35 & Ne\,\textsc{x} $1s\rightarrow2p$ & 12.14 & $-0.067\pm0.004$ \\
\enddata
\tablenotetext{a}{Centroid rest wavelength of an absorption line in \AA, with typical uncertainty of $\Delta\lambda=0.04$\AA.}
\tablenotetext{b}{Probable line ID and known laboratory wavelength from www.nist.gov.}
\tablenotetext{c}{Derived 
blue-shift of absorption line, in units of $c$. The majority of lines are systematically blueshifted with a mean velocity of $-0.062\pm0.001c$.
$^*$ denotes a possible higher velocity component.}
\label{tab:absorption-lines}
\end{deluxetable}


In addition to the narrow blueshifted absorption lines, a broad absorption trough is also clearly present, 
centered at $\sim$15.3\,\AA\ in the quasar rest frame (Figure~\ref{fig:mean_ratio}; upper panel). 
This is most likely associated with the the Fe M-shell 
Unresolved Transition Array (UTA), which originates from a blend of $2p\rightarrow3d$ transitions 
from low-ionization Fe (Fe\,\textsc{i-xvii}; \citealt{BeharSakoKahn01}). This broad feature has been 
observed in the warm absorbers in many type I AGN; e.g. IRAS\,13349+2438 \citep{Sako01}, NGC\,3783 \citep{Kaspi02, Krongold03}, NGC\,5548 \citep{Kaastra02}, Mrk\,509 \citep{Pounds01, Detmers11}, NGC\,7469 \citep{Blustin07}, Mrk\,841 \citep{Longinotti10}, IC\,4239A \citep{Steenbrugge05}, NGC\,3516 \citep{HolczerBehar12}, MCG-6-30-15 \citep{Lee01, Turner04}, NGC\,4051 \citep{Pounds04}, Mrk\,279 \citep{Costantini07}. It occurs typically between 16--17\,\AA, 
depending on the exact ionization of the absorber. 

\begin{figure*}
\begin{center}
\rotatebox{-90}{\includegraphics[height=15cm]{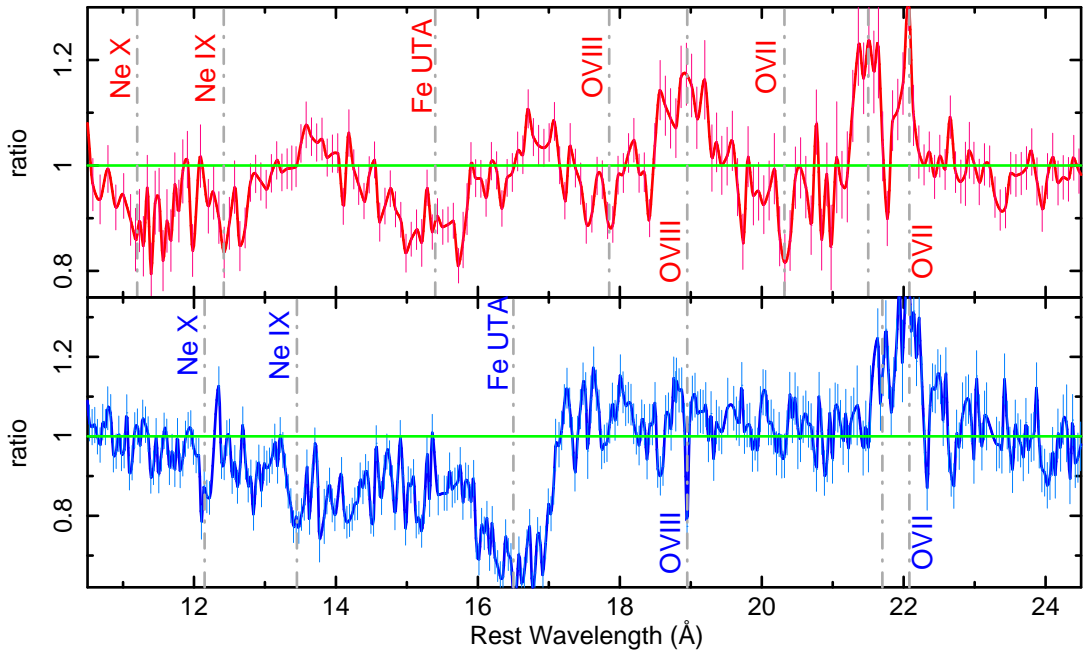}}
\end{center}
\caption{Comparison between the mean RGS spectrum of PG\,1211+143 (upper panel, red points) and the nearby ($z=0.064$) radio-quiet QSO, MR\,2251-178 (lower panel, blue points), plotted over the 10-25\,\AA\ range. MR\,2551-178 exhibits a classic slow warm absorber, with little or no blueshift (see \citealt{Reeves13}), where the strongest absorption/emission features are marked with dashed lines in their rest-frame wavelength. In contrast, the strong absorption features from PG\,1211+143, for instance from O\,\textsc{vii}, O\,\textsc{viii}, Ne\,\textsc{ix} and Ne\,\textsc{x} and the Fe M-shell UTA, are all systematically blueshifted compared to MR\,2251-178.}
\label{fig:ratio_vs_mr2251}
\end{figure*}

In Figure~\ref{fig:ratio_vs_mr2251}, we compare the spectrum of PG\,1211+143 with another nearby QSO of similar luminosity, MR\,2251-178 (at $z=0.064$), which is well known for its soft X-ray warm absorber (e.g. \citealt{Halpern84, Kaspi04, Reeves13}). The RGS data for the latter was taken from a 
long $\sim$300\,ks {\it XMM-Newton} RGS observation in 2011, presented in \citet{Reeves13}.  In MR\,2251-178 the absorber outflow velocity is low ($<1\,000$\,km\,s$^{-1}$) providing an interesting contrast to PG\,1211+143.

The rest-frame 
comparison between the two spectra clearly shows that the broad UTA is present towards both quasars, 
but it is blueshifted in PG\,1211+143 (centered at 15.3\,\AA) compared to what is measured in 
MR\,2251-178 (where the UTA is centered close to 16.5\,\AA). 
While some variation in the UTA 
position can be caused by variable ionization, due to the relative contribution of the different Fe ions towards the UTA \citep{BeharSakoKahn01}, as we will show later in Section~\ref{sec:photoionization_modelling}, the relative blueshift of the UTA in PG\,1211+143
is consistent with an outflow velocity of $-0.06c$.
In addition,  it is also apparent when compared to MR\,2251-178, that the principle narrow absorption lines are systematically blueshifted in PG\,1211+143. For instance, the Ne\,\textsc{ix} and Ne\,\textsc{x} $1s\rightarrow 
2p$ resonance absorption lines are detected very close to their expected rest wavelengths in 
MR\,2251-178 (at 13.4\,\AA\ and 12.1\,\AA, respectively) with little blueshift, while in PG\,1211+143 these lines are 
strongly blueshifted (12.4\,\AA\ and 11.3\,\AA, respectively). A similar case is seen for the O\,\textsc{viii} Ly$\alpha$ 
line, which is observed very close to its known wavelength near $\sim19$\,\AA\ in MR\,2251-178, but is blue-shifted (at 17.9\,\AA) in PG\,1211+143, with respect to the strong emission line that is observed at 19\,\AA. Likewise, the O\,\textsc{vii} $1s\rightarrow 2p$ absorption line (at 20.3\,\AA) is blueshifted with respect 
to the corresponding resonance emission line in MR\,2251-178 at 21.6\,\AA.

Strong emission lines are also seen in PG\,1211+143, which are apparent in Figures~\ref{fig:mean_ratio} and~\ref{fig:ratio_vs_mr2251}, 
particularly from O\,\textsc{vii} and O\,\textsc{viii}. The parameters of the significant emission lines fitted 
with Gaussian profiles are listed in Table~\ref{tab:emission-lines}. Most notable is the O\,\textsc{vii} triplet, which is 
resolved into a forbidden component (at 22.1\,\AA) and a resonance line ($21.46\pm0.06$\,\AA), the latter of which is mildly blueshifted with respect to its expected rest wavelength (at 21.60\,\AA). 
Generally, the emission line profiles are all significantly broadened with respect to the RGS resolution, 
the exception being the O\,\textsc{vii} forbidden line, which is relatively narrow (FWHM $\sim1\,200$\,km\,s$^{-1}$). For the other emission lines, the typical FWHM velocity widths are in the range from $\sim$5\,000-10\,000\,km\,s$^{-1}$; indeed, one of the strongest emission lines is from O\,\textsc{viii} Ly$\alpha$ at 19\,\AA, which has a FWHM of $\sim$11\,000\,km\,s$^{-1}$ (see Table~\ref{tab:emission-lines}). These widths are much broader than the typical optical permitted emission lines seen from PG\,1211+143, where the H$\beta$ width is relatively narrow with a FWHM of 1\,860\,km\,s$^{-1}$ \citep{BorosonGreen92}. This suggests the soft X-ray line emission could be associated with gas coincident with an innermost X-ray component of the Broad Line Region (BLR), or that it arises from re-emission associated with a wide angle wind. We discuss the latter possibility further later (see Section~\ref{sec:emission_components}).

\begin{deluxetable*}{lcccccc}
\tablecaption{Soft X-ray Emission Lines in PG\,1211+143 RGS spectrum.}
\tablewidth{0pt}
\tablehead{
\colhead{Line ID} & \colhead{$\lambda_{\rm quasar}^{a}$} & \colhead{Flux$^{b}$} & \colhead{EW$^{c}$} 
& \colhead{$\sigma_{\rm v}$$^{d}$} & \colhead{FWHM$^{e}$} 
& \colhead{$\Delta \chi^{2}$$^{f}$}}
\startdata
N\,\textsc{vii} Ly$\alpha$  & $24.98\pm0.16$ ($496.2^{+3.2}_{-3.2}$) & $2.7^{+1.6}_{-1.2}$ & $1.0^{+0.6}_{-0.4}$ & 
$3100^{+1800}_{-1500}$ & $7300^{+4200}_{-3500}$ & 14.8 \\
O\,\textsc{vii} (f) & $22.07\pm0.03$ ($561.8^{+0.8}_{-0.6}$) & $2.4^{+0.8}_{-0.6}$ & $1.4^{+0.5}_{-0.4}$ & 
$530^{+500}_{-270}$ & $1240^{+1200}_{-630}$ & 26.5 \\
O\,\textsc{vii} (r) & $21.46\pm0.11$ ($577.8^{+3.2}_{-3.0}$) & $2.4^{+1.0}_{-1.0}$ & $1.5^{+0.6}_{-0.6}$ & 
$2230^{+1300}_{-780}$ & $5250^{+3050}_{-1830}$ & 15.4 \\
O\,\textsc{viii} Ly$\alpha$ & $18.96\pm0.08$ ($653.9^{+2.9}_{-2.9}$) & $5.0^{+1.2}_{-1.0}$ & $4.9^{+1.2}_{-1.0}$ & 
$4700^{+1400}_{-960}$ & $11100^{+3700}_{-2300}$ & 63.2 \\
O\,\textsc{viii} RRC$^g$ & $16.84\pm0.09$ ($736.0^{+4.0}_{-4.0}$) & $1.9^{+0.6}_{-0.5}$ & $2.8^{+0.9}_{-0.7}$ & 
$-$ & $-$ & 39.1 \\
Ne\,\textsc{ix}$^{h}$ (f) & $13.68\pm0.11$ ($906.0^{+7.0}_{-7.0}$) & $1.0^{+0.3}_{-0.3}$ & $2.7^{+1.0}_{-1.0}$ & 
$4000^{+2000}_{-1700}$ & $9400^{+4600}_{-4000}$ & 21.3\\
\enddata
\tablenotetext{a}{Rest wavelength in units of \AA\ in quasar rest frame. Rest energy in eV given in parenthesis.}
\tablenotetext{b}{Photon flux in units of $\times10^{-5}$\,photons\,cm$^{-2}$\,s$^{-1}$.}
\tablenotetext{c}{Equivalent width in quasar rest frame in units of eV.}
\tablenotetext{d}{$1 \sigma$ velocity width in units of km\,s$^{-1}$.}
\tablenotetext{e}{FWHM velocity width in units of km\,s$^{-1}$.}
\tablenotetext{f}{Improvement in $\Delta \chi^{2}$ upon adding line to model.}
\tablenotetext{g}{Temperature of RRC constrained to $kT=18^{+14}_{-6}$\,eV.}
\tablenotetext{h}{Note that the broad Ne\,\textsc{ix} line may also be blended with a contribution from a
O\,\textsc{viii} RRC. }  
\label{tab:emission-lines}
\end{deluxetable*}

We also fitted some of the absorption lines with Gaussian profiles, using the 
$1s\rightarrow 2p$ lines from N\,\textsc{vii}, O\,\textsc{vii} and O\,\textsc{viii}, which are relatively isolated line profiles. For these three lines we obtain a mean Gaussian width of $\sigma=800^{+350}_{-240}$\,km\,s$^{-1}$ (or FWHM $\sim1\,800$\,km\,s$^{-1}$). In contrast, the $1\sigma$ velocity resolution of the RGS is 
$\sigma\sim4 00$\,km\,s$^{-1}$ at the position of the O\,\textsc{vii} absorption line. Thus, unlike for the emission lines, the absorption lines 
appear to be marginally resolved compared to the instrumental resolution.

\subsection{Photoionization Modelling} \label{sec:photoionization_modelling}

Next, we constructed a photoionized absorption model to account for the mean 2014 RGS spectrum 
of PG\,1211+143. 
The continuum form with 2 power laws described above was adopted.   
The phenomenological form of the model is:- \\
$\textsc{tbabs} \times [\textsc{xstar}_{\textsc{abs}} \times (\textsc{pow}_{\textsc{hard}} + \textsc{bbody}) + \textsc{pow}_{\textsc{soft}} + \textsc{xstar}_{\textsc{emission}}].$ 

\noindent Here, $\textsc{xstar}_{\textsc{abs}}$ represents the multiplicative photoionized absorption grids of models generated by \textsc{xstar} \citep{Kallman96} that are applied to the continuum, while $\textsc{xstar}_{\textsc{emission}}$ 
represents the additive photoionized emission tables generated by \textsc{xstar} in order to self-consistently fit the line emission (see below). Only one of the power law continuum components ($\textsc{pow}_{\textsc{hard}}$) is absorbed by the photoionized absorbers, 
while the other power law ($\textsc{pow}_{\textsc{soft}}$) emerges unattenuated by the 
photoionized absorbers. 
The neutral Galactic absorption component (\textsc{tbabs}) absorbs all of the emission 
components. The \textsc{bbody} component is initially included to account for any additional soft excess that may be required, as may originate from the high energy Wien tail of any disc emission. 
However such a component was not required over the wavelength range analysed in the RGS spectra (up to 30\,\AA\ observed frame) and thus was subsequently omitted from the analysis.

The \textsc{xstar} grids of models are identical to the ones that we generated in a previous analysis of 
PG\,1211+143 (e.g. \citealt{Pounds16a, Pounds16b}). These were calculated with the spectral energy distribution (SED) derived for PG\,1211+143 
from the simultaneous {\it XMM-Newton} Optical Monitor photometry and the mean 2014 EPIC-pn spectrum, 
which was parametrized by a double-broken-power law model; see \citet{Lobban16a} for details. 
The total 1--1000\,Rydberg ionizing luminosity from this SED was found to be $4\times10^{45}$\,erg\,s$^{-1}$, which was input into the \textsc{xstar} models.

\begin{deluxetable*}{lcccc}
\tablecaption{Properties of the photoionized absorber/emission components fitted to the mean RGS spectrum.}
\tablewidth{0pt}
\tablehead{
\colhead{Parameters} & \colhead{Zone\,1a} & \colhead{Zone\,1b} & \colhead{Zone\,2}& \colhead{Zone\,3}}
\startdata
Absorption:-\\
$N_{\rm H}$ ($\times$10$^{21}$\,cm$^{-2}$) & $1.40^{+0.42}_{-0.36}$ & $10.0^{+6.0}_{-4.8}$ & 
$0.89^{+0.35}_{-0.27}$ & $1.6^{+0.8}_{-0.6}$\\
$\log\xi$ & $1.32^{+0.15}_{-0.20}$ & $3.37^{+0.08}_{-0.12}$ & $2.03^{+0.17}_{-0.11}$ & $2.51^{+0.12}_{-0.20}$\\
$v/c$  & $-0.062\pm0.001$ & $-0.059\pm0.002$ & $-0.077\pm0.001$ & $-0.187\pm0.002$\\
$\Delta\chi^{2}$$^{a}$ & 121.9 & 28.3 & 21.1 & 18.5 \\
$P_{\rm N}$$^{b}$ & $5.1\times10^{-25}$ & $2.2\times10^{-5}$ & $5.4\times10^{-4}$ & $9.6\times10^{-4}$ \\
\hline\\
Emission:-\\
$N_{\rm H}$ ($\times$10$^{21}$\,cm$^{-2}$) & 1.4$^{t}$ & 10$^{t}$ \\
$\log\xi$ & $2.0^{+0.4}_{-0.3}$ & $3.4^{t}$ \\
$v/c$  & $<0.002$ & $-0.017\pm0.006$ \\
$\kappa_{\rm xstar}$$^{c}$ & $0.70\pm0.25$ & $1.7\pm0.7$\\
$f_{\rm cov}=\Omega/4\pi$$^{d}$ & $0.20\pm0.06$ & $0.46\pm0.20$ \\
$\Delta\chi^{2}$$^{a}$ & 52.4 & 21.7\\
$P_{\rm N}$$^{b}$ & $2.4\times10^{-9}$ & $3.1\times10^{-4}$ & \\
\hline
\enddata
\tablenotetext{a}{Improvement in fit statistic from adding the absorption or emission zone to the model.}
\tablenotetext{b}{Null probability of adding the \textsc{xstar} zone, calculated via an F-test.}
\tablenotetext{c}{Measured normalization of \textsc{xstar} emission component, where $\kappa_{\rm xstar}= f_{\rm cov} L_{38}/D_{\rm kpc}^2$, where $f_{\rm cov}$ is the covering fraction ($f=\Omega/4\pi$), $L_{38}$ is the 1--1000\,Ryd 
ionizing luminosity in units of $10^{38}$\,erg\,s$^{-1}$ and $D_{\rm kpc}$ is the distance to PG\,1211+143 
in units of kpc.}
\tablenotetext{d}{Global covering fraction of the \textsc{xstar} emission component, derived from the 
measured normalization of the emitter. See text for details.}
\tablenotetext{f}{Denotes parameter is fixed.}
\label{tab:xstar}
\end{deluxetable*}

Both multiplicative absorption tables as well as additive emission tables (in the forwards direction) 
were generated to account for both the ionized absorption and emission lines observed in the 
soft X-ray spectrum. Grids were run for 3 different turbulence velocities\footnote{The turbulence velocity width is defined as $b=\sqrt{2} \sigma = {\rm FWHM}/(2\sqrt{\ln 2})$.}, with $b=300, 1\,000, 3\,000$\,km\,s$^{-1}$. Note that the lower two turbulences correspond to velocity widths which well match the relatively narrow 
absorption lines, where the mean Gaussian width is $\sigma=800^{+350}_{-240}$\,km\,s$^{-1}$ (or FWHM $\sim1\,800$\,km\,s$^{-1}$) and give statistically equivalent fits. 
The broader $b=3\,000$\,km\,s$^{-1}$ grids were used to fit the photoionized emission, as the emission lines appears substantially broadened, as described in Section~\ref{sec:soft_x-ray_absorption_emission}. 
Each grid was also generated with three different Fe abundances relative to the Solar values of \citet{GrevesseSauval98}, consisting of $\times1$, $\times3$ and $\times 5$ Solar for Fe. As discussed in 
\citet{Pounds16b}, a higher abundance of Fe is driven by the requirement to fit the strong UTA 
absorption. Indeed, a worse fit was obtained using a Solar abundance of Fe, whereas 
statistically equivalent fits were obtained for $\times 3$ and $\times 5$ abundances of Fe; thus, grids 
with $\times3$ Fe abundance were subsequently adopted in all of the fits. 


Prior to modeling the photoionized emission and absorption, the overall 
fit statistic to the baseline double power law continuum model 
is extremely poor, with a reduced chi-squared of $\chi_{\nu}^2=956.3/595=1.61$ and a 
corresponding null hypothesis probability for the model to be acceptable of $p=2.7\times10^{-19}$. 
Thus, such a simple model, with no ionized emission or absorption, is rejected by the data 
at a high level of significance, which is apparent from the data/model residuals present in Figure~\ref{fig:mean_ratio}.

\begin{figure*}
\begin{center}
\rotatebox{-90}{\includegraphics[height=16cm]{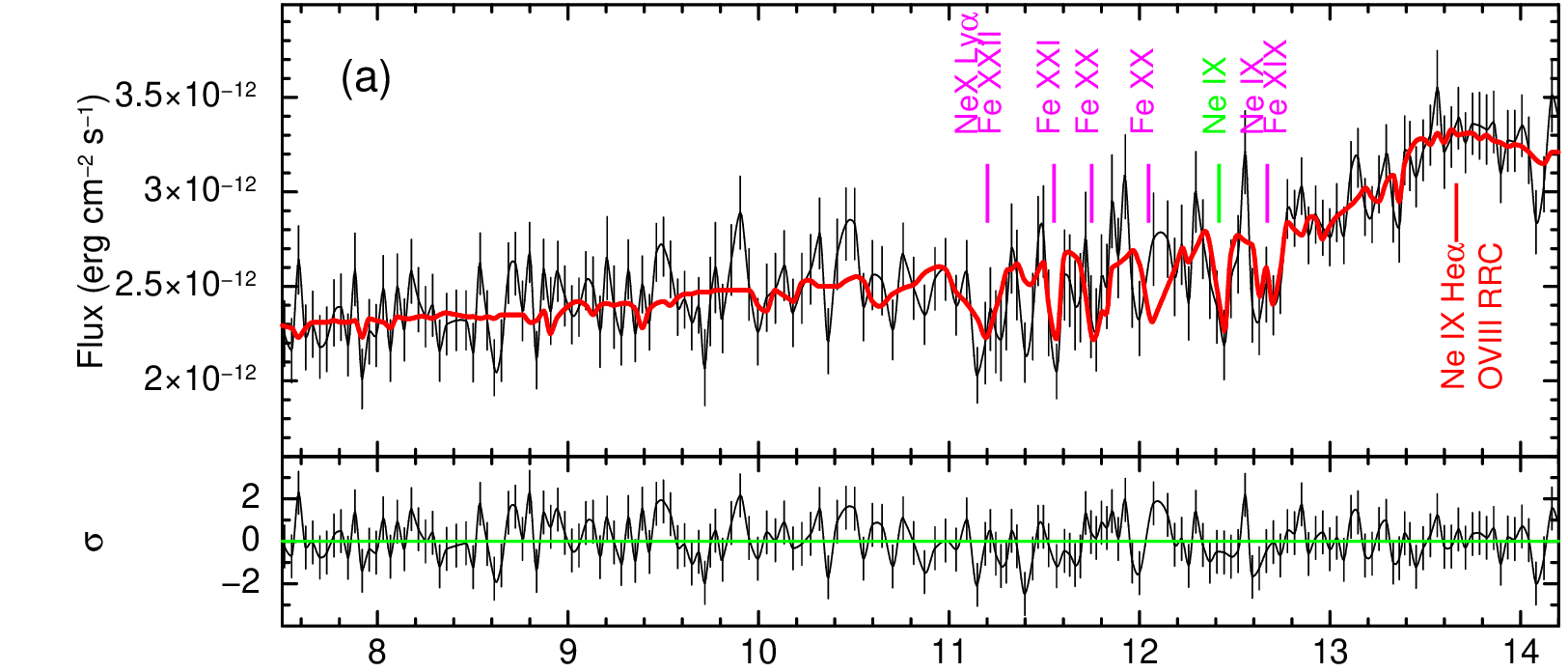}}
\rotatebox{-90}{\includegraphics[height=16cm]{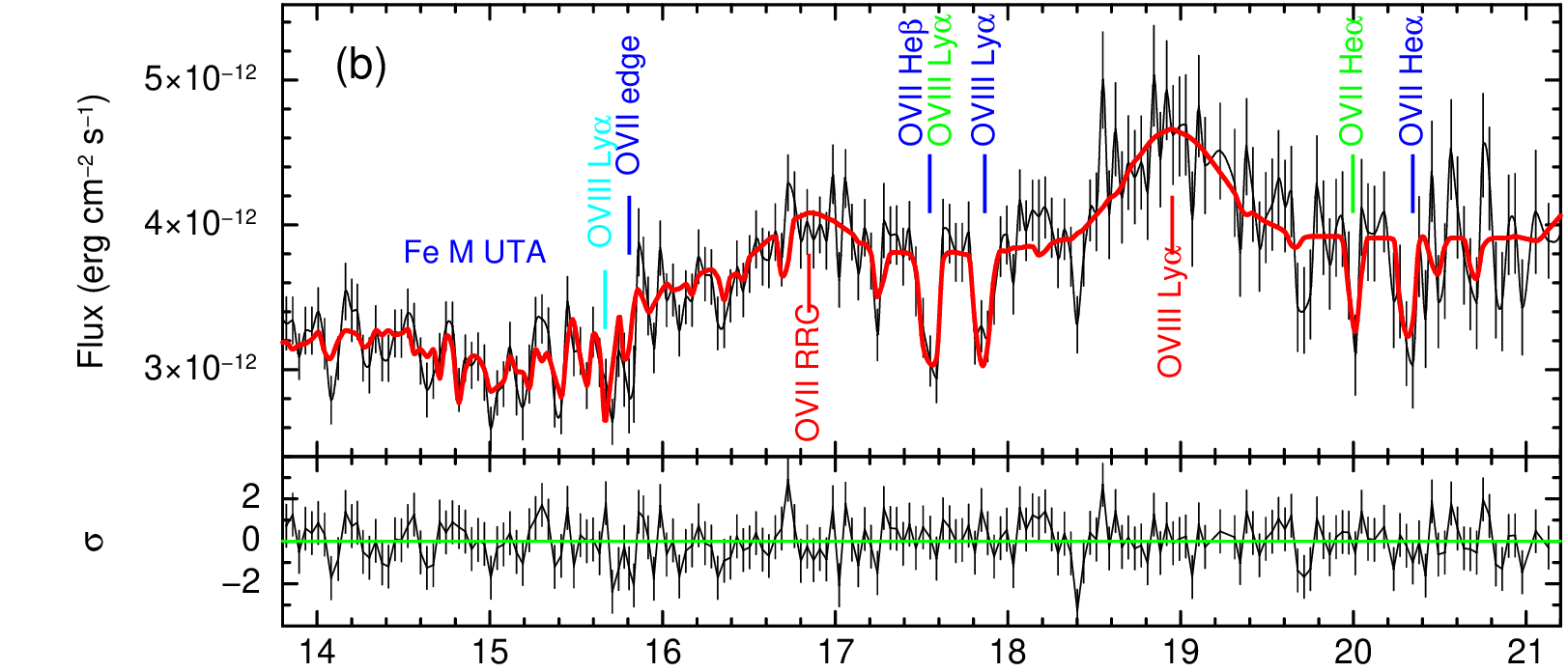}}
\rotatebox{-90}{\includegraphics[height=16cm]{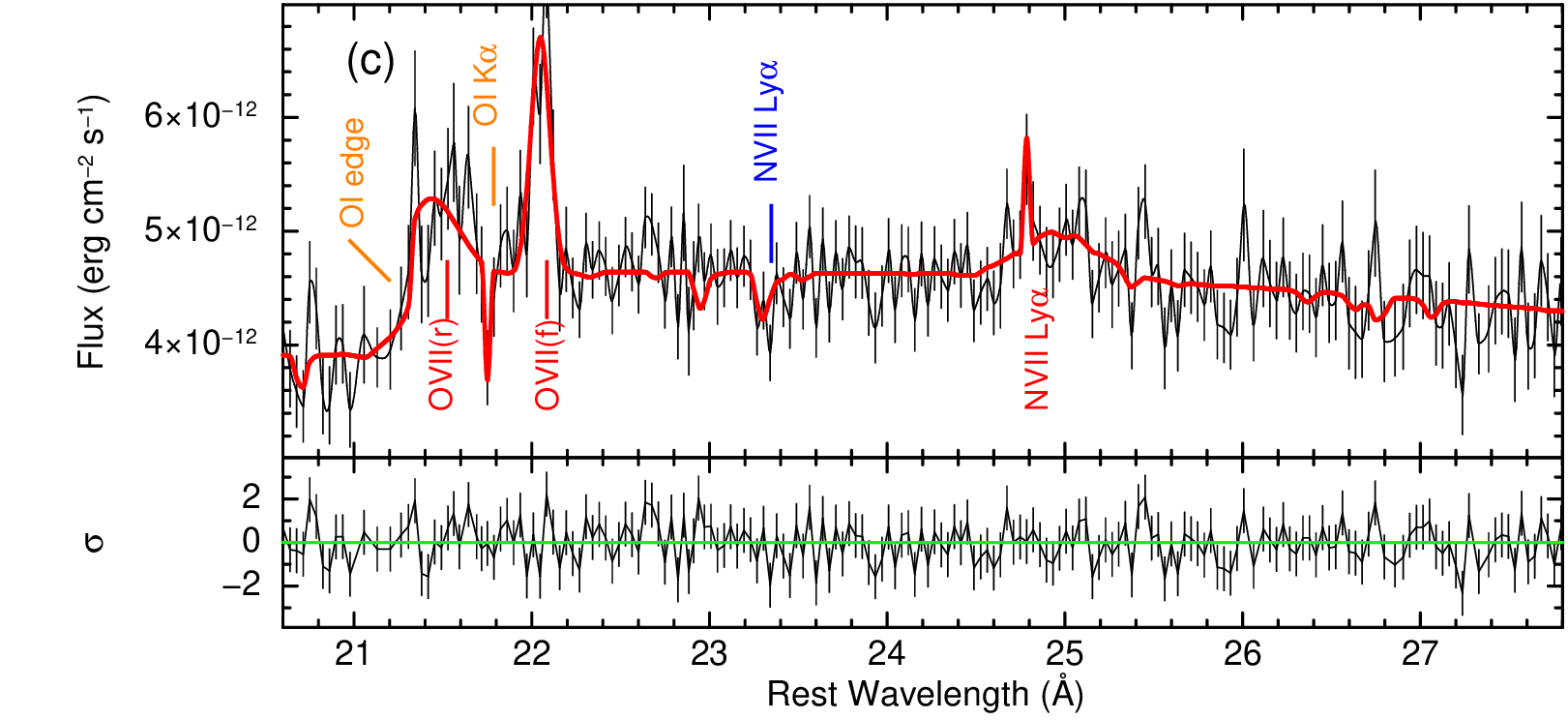}}
\end{center}
\caption{ The best-fitting \textsc{xstar} model (red line), fitted to the mean RGS spectrum of PG\,1211+143; panel (a) shows the iron L-shell band, panel (b) the O\,\textsc{vii-viii} and Fe M-shell band and panel (c) the O\,\textsc{vii} 
triplet and N K-shell band. The spectrum is fitted down to 27.8\,\AA\ in the 
quasar rest frame (30\,\AA\ in the observed frame). 
The strongest atomic features are marked on the 
panels and are color coded to represent the different absorption and emission zones listed in Table~\ref{tab:xstar}. The blue-labels correspond to absorber 
zone 1a (low ionization, Fe M-shell UTA and He/H-like N/O) and magneta labels correspond to zone 1b (high ionization, Fe L-shell and He/H-like Ne), both at an outflow velocity of $-0.06c$. Green labels correspond to zone 2 with an outflow velocity of $-0.08c$ (seen from O\,\textsc{vii}, O\,\textsc{viii} and Ne\,\textsc{ix}) and the cyan label corresponds to zone 3 ($-0.18c$), although the latter ID is uncertain. Red labels correspond to the broadened emission lines 
present in the spectrum, which in particular include 
the forbidden and resonance components of O\,\textsc{vii} and associated RRC emission at $\sim$17\,\AA, as well as a strong, broad (FWHM $\sim 10\,000$\,km\,s$^{-1}$) emission line from O\,\textsc{viii} Ly$\alpha$. 
The orange labels mark the O\,\textsc{i} features produced by the Galactic 
($z=0$) neutral absorber. Overall the model reproduces the observed features in 
the spectrum and most of the remaining residuals agree within $\pm2\sigma$.}
\label{fig:mean_xstar}
\end{figure*}

\subsection{Absorption Components} \label{sec:absorption_components}

In order to obtain an acceptable fit to the ionized absorption present in the spectrum, successive absorption zones were then applied to the spectrum, until the addition of further absorption grids no longer 
improved the fit at the 99\,per cent confidence level. Emission was also included by additive photoionized 
emission grids and is described further below (Section~\ref{sec:emission_components}). The parameters of the absorption zones are summarized 
in Table~\ref{tab:xstar}, while the overall best-fit to the \textsc{xstar} model is plotted in Figure~\ref{fig:mean_xstar}.
Here, four absorbers (zones 1a, 1b, 2 and 3) were required at $>99.99$\% confidence and these are described further below. 

\begin{figure}
\begin{center}
\rotatebox{-90}{\includegraphics[height=8.5cm]{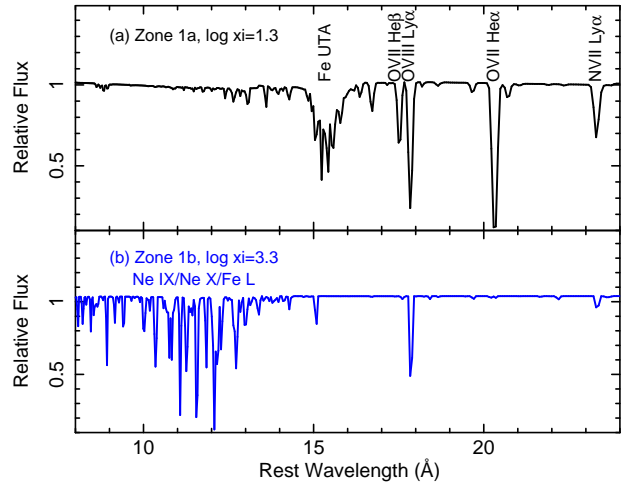}}
\end{center}
\caption{Relative transmission through the two main ionization zones of the absorbers.  The upper panel shows 
the low-ionization zone\,1a ($\log\xi=1.3$) and the lower panel the high-ionization zone\,1b ($\log\xi=3.3$). 
Both zones are outflowing with $v/c=-0.06$ (see Table~\ref{tab:xstar} for details). The lower-ionization zone\,1a reproduces 
the strong blueshifted Fe M-shell UTA, as well as the He/H-like lines of N and O. The high-ionization zone\,1b 
reproduces the strong series of Fe L-shell lines between 10--12\AA, as well as absorption from 
Ne\,\textsc{ix-x} and O\,\textsc{viii}.}
\label{fig:model}
\end{figure}

As noted earlier from the Gaussian line analysis, 
most of the absorption is associated with gas outflowing with $v/c=-0.06$, 
which are parametrized by zones 1a and 1b. Adding these zones of absorption against the baseline continuum lead to corresponding improvements 
in fit statistic of $\Delta\chi^2 = 121.9$ and $\Delta\chi^2 = 28.3$, respectively. 
The relative contributions of these two absorption zones are also shown in Figure~\ref{fig:model}. 
Zone\,1a is a lower-ionization zone, with $N_{\rm H}=1.4\pm0.4\times10^{21}$\,cm$^{-2}$, 
an ionization parameter\footnote{The ionization parameter is defined as $\xi = L_{\rm ion}/nR^2$ \citep{TarterTuckerSalpeter69}, where $L_{\rm ion}$ is the 1--1000 Rydberg ionizing luminosity, $n$ is the electron density and $R$ is the distance of the ionizing source from the absorbing clouds. The units of $\xi$ are erg\,cm\,s$^{-1}$.}
of $\log\xi=1.32^{+0.15}_{-0.20}$ and an outflow velocity of $v/c=-0.062\pm0.001$. 
It is the most significant zone, carrying most of the absorber opacity, including the Fe M-shell UTA 
and the He/H-like lines of N and O; the former of which also appears to 
be associated with gas at a velocity of $\sim -0.06c$ rather than a slow warm-absorber component.
Indeed these features (marked by blue labels) 
are well modeled in the best fit spectrum shown in Figure~\ref{fig:mean_xstar}, 
including the broad Fe M-shell UTA at around 15\,\AA\, and the $1s\rightarrow2p$ resonance 
lines from O\,\textsc{viii} (17.9\AA), 
O\,\textsc{vii} (20.4\,\AA) and N\,\textsc{vii} (23.3\AA).
 
Zone\,1b represents the high-ionization phase of the $\sim -0.06c$ outflow, with a column density of 
$N_{\rm H}=1.0^{+0.6}_{-0.4}\times10^{22}$\,cm$^{-2}$, an ionization of $\log\xi=3.4\pm0.1$ and a 
similar outflow velocity of $v/c=-0.059\pm0.002$. This zone contributes towards the 
blend of absorption between $\sim 11-13$\,\AA\ (panel a, Figure~\ref{fig:mean_xstar}, magneta labels) due to higher-ionization L-shell Fe (Fe\,\textsc{xix-xxii} $2p\rightarrow3d$). 
This zone also contributes towards absorption from the $1s\rightarrow2p$ transitions 
from Ne\,\textsc{ix} (12.6\AA) and Ne\,\textsc{x} (11.2\AA)
as well towards the O\,\textsc{viii} Ly$\alpha$ absorption line at 17.9\,\AA. 

The addition of these two $-0.06c$ zones largely account for most of the ionized absorption present in the 
spectrum shown in Figure~\ref{fig:mean_xstar}. Nonetheless, there is some evidence for two possible higher-velocity zones. 
Zone\,2 is associated with gas with an outflow velocity of $v/c=-0.077\pm0.001$, with a column density 
of $N_{\rm H}=0.9\pm0.3\times10^{21}$\,cm$^{-2}$ and an ionization of $\log\xi=2.03^{+0.17}_{-0.11}$; 
its addition to the model led to a fit improvement of $\Delta \chi^2=21.1$. 
Its contribution towards the spectrum is more subtle, effectively adding a higher-velocity component bluewards of the O\,\textsc{vii} and O\,\textsc{viii} $1s\rightarrow 2p$ line profiles near to $\sim20.0$\,\AA\ and 
$\sim17.6$\,\AA\, respectively (see Figure~\ref{fig:mean_xstar}; green labels).
Note that the latter line may also be, in part, associated with O\,\textsc{vii} He$\beta$ ($1s\rightarrow 3p$) at $-0.06c$, 
as was discussed above (Section~\ref{sec:soft_x-ray_absorption_emission}) in the Gaussian line analysis.
Zone 2 may also contribute towards a $-0.08c$ component of Ne\,\textsc{ix} at 12.4\,\AA\ (Figure~\ref{fig:mean_xstar}, panel a).

A third high-velocity component (zone\,3), with $v/c=-0.187\pm0.002$, 
is formally required ($\Delta\chi^2=18.5$) and was noted by \citet{Pounds16b}, in addition to the above slower zones, in an earlier analysis of the mean spectrum. However, unlike for the $-0.06c$ and $-0.08c$ zones, the association of this absorption zone 
with discrete lines in the RGS spectrum is not easily apparent. 
Its main effect is to produce an absorption 
dip around 15.7\,\AA\ near the UTA due to blueshifted O\,\textsc{viii} Ly$\alpha$, (see Figure~\ref{fig:mean_xstar}, panel b, cyan label) and thus given its uncertain identification 
this additional fast zone is not considered further.

In addition to the above outflowing absorption zones, a neutral absorption component is required 
associated with our Galaxy at $z=0$. The best-fit column density is $N_{\rm H}=6.5^{+1.0}_{-1.2}\times10^{20}$\,cm$^{-2}$ and is about twice that predicted from the 21\,cm value of $N_{\rm H}=2.9\times10^{20}$\,cm$^{-2}$. 
The excess column density may be due to molecular hydrogen along the line of sight in our Galaxy, 
which would otherwise be missed in 21\,cm surveys, but would still be detected in the X-ray spectrum \citep{Willingale13}. 
Note that we also cannot rule out some contribution 
from neutral gas associated with the host galaxy of PG\,1211+143. 
The neutral absorber contributes towards the O\,\textsc{i} edge, observed at $z=0$ 
near 23.0\,\AA\ (or 21.4\,\AA\ 
in the QSO frame), as well as a weak O\,\textsc{i} K$\alpha$ absorption line observed at 23.5\,\AA\ (or 21.75\,\AA\ in the QSO frame). These neutral absorption 
features are apparent in Figure~\ref{fig:mean_xstar}, 
as marked by the orange labels in panel (a) in the O K-shell band.

Note that the best-fit absorption model (together with emission lines as modeled below) produces 
a final acceptable fit of $\chi_{\nu}^2=631.5/573=1.10$, while generally there are 
no other clear systematic residuals within the $\pm2\sigma$ level
present against the best fit shown in Figure~\ref{fig:mean_xstar}.
The only remaining residual at $>2\sigma$ against the best fit model lies at 18.4\,\AA, just bluewards 
of the O\,\textsc{viii} Ly$\alpha$ emission line (panel b, Figure~\ref{fig:mean_xstar}). 
If this absorption feature is associated to O\,\textsc{viii} Ly$\alpha$, then its outflow velocity is $-0.03c$, which is intermediate between the above fast outflow zones and a slower warm absorber. 
However, no such velocity component appears to be associated with any of the other ions 
and we note that one would typically expect to 
observe 1--2 features purely by statistical chance at the $3\sigma$ level over a 600 channel spectrum.

As a final consistency check, we also tested a Solar abundance absorber, by replacing the  
$\times3$ Fe abundance grid with the respective Solar one, for all of the zones. 
The overall fit statistic is acceptable, $\chi_{\nu}^2=650.7/573$, although it is still formally worse 
than the $\times3$ Solar case (by $\Delta\chi^2=19.2$). The difference in the fits are subtle, 
but the main difference arises from the O features being slightly overpredicted with respect to the UTA, 
with the high ionization iron L-shell features being somewhat underpredicted. 
However, overall the fit parameters are all consistent with those listed in Table\,4, 
with the only significant change being an increase in column of the low ionization 
zone 1a from $N_{\rm H}=1.4\pm0.4\times10^{21}$\,cm$^{-2}$ to $N_{\rm H}=2.5\pm0.6\times10^{21}$\,cm$^{-2}$
in order to reproduce the depth of the UTA.

\subsubsection{The Lack of a Warm Absorber} \label{sec:warm_absorber}
  
Importantly, there do not appear to be any absorption features with zero (or low) velocity shift 
in the spectrum, which could be associated to a low velocity warm absorber that are more commonly 
observed in Seyfert 1 galaxies. For instance, from inspecting Figure~\ref{fig:mean_xstar}, 
there appears to be no absorption residuals at wavelengths close to 
the expected transitions of N\,\textsc{vii} Ly$\alpha$ (at 24.78\,\AA), 
O\,\textsc{vii} $1s\rightarrow2p$ (21.60\,\AA), O\,\textsc{viii} Ly$\alpha$ (18.97\,\AA) or 
Ne\,\textsc{ix} $1s\rightarrow2p$ (13.45\,\AA), all of which are frequently detected in 
warm absorber spectra. Indeed the absorption features which are observed are strongly blueshifted 
with respect to their corresponding emission features, e.g. such as O\,\textsc{vii} He$\alpha$ or O\,\textsc{viii} Ly$\alpha$.

Nonetheless, we place upper-limits on the presence of a warm absorber in the PG\,1211+143 spectrum. 
We tested the presence of a warm absorber from adopting the same \textsc{xstar} absorption grid, 
but restricting the outflow velocity to within $\pm0.01c$ of the systemic velocity of PG\,1211+143. 
In this case, over three different ionization regimes, we find upper limits of $N_{\rm H}<2.5\times10^{19}$\,cm$^{-2}$, $N_{\rm H}<4.0\times10^{19}$\,cm$^{-2}$ and $N_{\rm H}<1.8\times10^{20}$\,cm$^{-2}$, for 
$\log\xi=1,2,3$ respectively.
These limits are more than an order of magnitude lower than the respective ionization zones for the fast absorber, 
ruling out the presence of a warm absorber in PG\,1211+143.

\subsection{Emission Components} \label{sec:emission_components}

In addition to the absorption zones, two zones of emission are required to fit the soft X-ray emission lines 
present in PG\,1211+143, where the strongest emission features are marked by red labels in Figure~\ref{fig:mean_xstar}.
Note that higher turbulence ($b=3\,000$\,km\,s$^{-1}$) grids were used 
to match the higher-velocity broadening of the emission features. The exception to this is the 
narrow O\,\textsc{vii} forbidden line at 22.1\,\AA\ which has instead been modelled with a separate 
narrow Gaussian profile. Given the lower velocity-width of this line (Table~\ref{tab:emission-lines}), it may be associated to more distant gas at pc-scales or greater. A similar (but weak) narrow component of 
emission may also be associated to N\,\textsc{vii} Ly$\alpha$ at 24.8\,\AA.

The two zones correspond to low- and high-ionization gas, with ionization parameters of $\log \xi=2.0$ and 
$\log \xi=3.4$, respectively (see Table~\ref{tab:xstar} for details of the emission zones). These zones may be ascribed to 
the re-emission from the corresponding absorption zones 1a and 1b, with similar ionization parameters, while 
the column density of the emission zones has been tied to the corresponding absorption zone for simplicity.
The two emission zones are both significant ($\Delta\chi^2=52.4, 21.7$ for the low- and high-ionization zones, respectively) and are able to reproduce the soft X-ray emission features, such as the broad O\,\textsc{viii} Ly$\alpha$ line (see Figure~\ref{fig:mean_xstar}). The net velocity of the lower-ionization emission zone is consistent with zero (within $v/c<0.002c$), while the high-ionization zone prefers a slight blueshift, with $v/c=-0.017\pm0.006$. 
Note the high-ionization zone predominately contributes towards a `blue wing' of the 
broad O\,\textsc{viii} Ly$\alpha$ emission line, as well some broad Ne\,\textsc{ix} emission.
One possibility is that the soft X-ray lines originate from the re-emission from a wide-angle outflow 
\citep{PoundsReeves09, Nardini15}. In this case, the breadth of the emission lines arises 
from the integration over a range of angles, from both the blue and redshifted sides of the wind, 
with typical velocity widths observed here up to 10\,000\,km\,s$^{-1}$ in the case of the 
O\,\textsc{viii} Ly$\alpha$ line.

\subsubsection{The Emitter Covering Fraction} \label{sec:covering_fraction}

In this framework, the luminosity of the soft X-ray line emission can also be used to calculate 
the global covering factor of the outflowing gas (see \citealt{Reeves16} for a similar calculation). 
From the photoionization modelling, 
the normalization (or flux), $\kappa$, of each of the emission components is defined 
by {\sc xstar} in terms of:
\begin{equation}
\kappa = f_{\rm cov}\frac{L_{38}}{D_{\rm kpc}^2},
\end{equation}
where $L_{38}$ is the 1--1000\,Rydberg ionizing luminosity in units of $10^{38}$\,erg\,s$^{-1}$ and 
$D_{\rm kpc}$ is the distance to the quasar in kpc. Here, $f_{\rm cov}$ is the covering fraction 
of the gas, with respect to the total solid angle, where $f_{\rm cov} = \Omega / 4\pi$ and thus $f_{\rm cov}=1$ for a spherical shell of gas. 
Thus, by comparing the predicted normalization ($\kappa$) for 
a fully-covering shell of gas (of column density, $N_{\rm H}$), 
illuminated by a luminosity, $L$, versus the observed 
normalization ($\kappa_{\rm xstar}$), determined from the photoionization modelling, 
the covering fraction of the gas can be estimated. 

For PG\,1211+143, with $L=4\times10^{45}$\,erg\,s$^{-1}$, at a distance of $D=331$\,Mpc, and for a spherical shell, then the expected {\sc xstar} normalization is $\kappa=3.7\times10^{-4}$. Compared to 
the observed normalization factors reported in Table~\ref{tab:rgs_parameters}, the covering fraction of 
lower-ionization zone of emitting gas is $f_{\rm cov}=0.20\pm0.06$ (for a column density of $N_{\rm H}=10^{21}$\,cm$^{-2}$), while the high-ionization zone has a covering fraction of $f_{\rm cov}=0.46\pm0.20$ (for $N_{\rm H}=10^{22}$\,cm$^{-2}$). 
This is consistent with the soft X-ray emitting gas covering a substantial fraction of 
$4\pi$\,steradian. This also supports the earlier interpretation of \citet{PoundsReeves09}, based on geometrical constraints from the broad emission-line profiles, of a wide-angle outflow in PG\,1211+143.

\subsection{The Continuum Parameters}

With all the absorption and emission components included in the model, the continuum is well described 
by a soft power law component with a photon index of $\Gamma=3.44^{+0.14}_{-0.17}$ and 
a hard power law of $\Gamma=1.60\pm0.15$.
The continuum parameters are summarized in Table~\ref{tab:continuum}.
Note these continuum parameters are broadly consistent with those obtained from a broad-band 
{\it XMM-Newton} and {\it NuSTAR} analysis of PG\,1211+143 \citep{Lobban16b}.
In the final fit, the blackbody component is not required as noted earlier and the continuum curvature is well described by the combination of the two power laws. 
For the above model construction, the absorption zones are applied against the hard power law, whereas 
the soft power law is unattenuated. If the opposite is assumed to be the case (where the soft power law 
is absorbed), then the fit statistic obtained is considerably worse ($\Delta\chi_{\nu}^2=57.7$), where in this case the shorter wavelength features such as the Fe M-shell UTA and the 1\,keV absorption due to Fe L are under-predicted, compared to the longer wavelength features such as the O band absorption, 
which are then over-predicted.

\begin{deluxetable}{lcc}
\tabletypesize{\small}
\tablecaption{Best-fit continuum parameters for the mean PG\,1211+143 spectrum.}
\tablewidth{0pt}
\tablehead{
\colhead{Component} & \colhead{Parameter} & \colhead{Value}}
\startdata
Soft Powerlaw & $\Gamma$ & $3.44^{+0.14}_{-0.17}$\\
 & $N_{PL}$$^{a}$ & $1.36^{+0.12}_{-0.09}$\\
\hline
Hard Powerlaw & $\Gamma$ & $1.60\pm0.15$\\
& $N_{PL}$$^{a}$ & $0.74^{+0.06}_{-0.07}$ \\
\hline
Galactic Column & $N_{\rm H}$$^{b}$ & $6.5^{+1.0}_{-1.2}$\\
\enddata
\tablenotetext{a}{Normalization of the powerlaw component, in units of flux at 1\,keV ($\times10^{-3}$\,photons\,cm$^{-2}$\,s$^{-1}$\,keV$^{-1}$).}
\tablenotetext{b}{Galactic column density in units of $\times10^{20}$\,cm$^{-2}$.}
\label{tab:continuum}
\end{deluxetable}

While the model construction is purely phenomenological, the steep power law seems likely to have a 
physical basis. 
In an analysis of the spectral variability between earlier (2001, 2004, 2007) {\it XMM-Newton} spectra, \citet{PoundsReeves09} noted that most of the spectral variability can be accounted for by the steep 
$\Gamma\sim3$ power law and that this component, as seen from the difference spectra between observations, appeared to carry 
little imprint of the ionized absorber in the soft band. Thus, when the spectrum brightened and the soft excess became stronger, the imprint of the absorber became weaker in these archival data. In a variability 
analysis of the current 2014 {\it XMM-Newton} datasets, \citet{Lobban16a} also showed that the long-term (inter-orbit) 
variability was also dominated by the steep power law component and when this component is 
low, the spectrum appears more absorbed. Furthermore \citet{Lobban16a} also noted inter-orbit changes in the 
strength of the UTA absorption.  The inter-orbit changes of the soft absorber in the RGS 
spectra are analyzed next in Section~\ref{sec:outflow_variability}, while some possible physical interpretations of the absorber variability are discussed in Section~\ref{sec:discussion}.

\section{Outflow Variability} \label{sec:outflow_variability}

The spectral model from the mean RGS spectrum was then 
used as a template to study the inter-orbit variability of the 
soft X-ray spectrum from \textsc{rev}\,2652 to \textsc{rev}\,2670. 
Indeed, as shown by \citet{Lobban16a}, the spectral variability 
of PG\,1211+143 is dominated by changes in the soft X-ray band. 
 
\begin{figure*}
\begin{center}
\rotatebox{0}{\includegraphics[width=15cm]{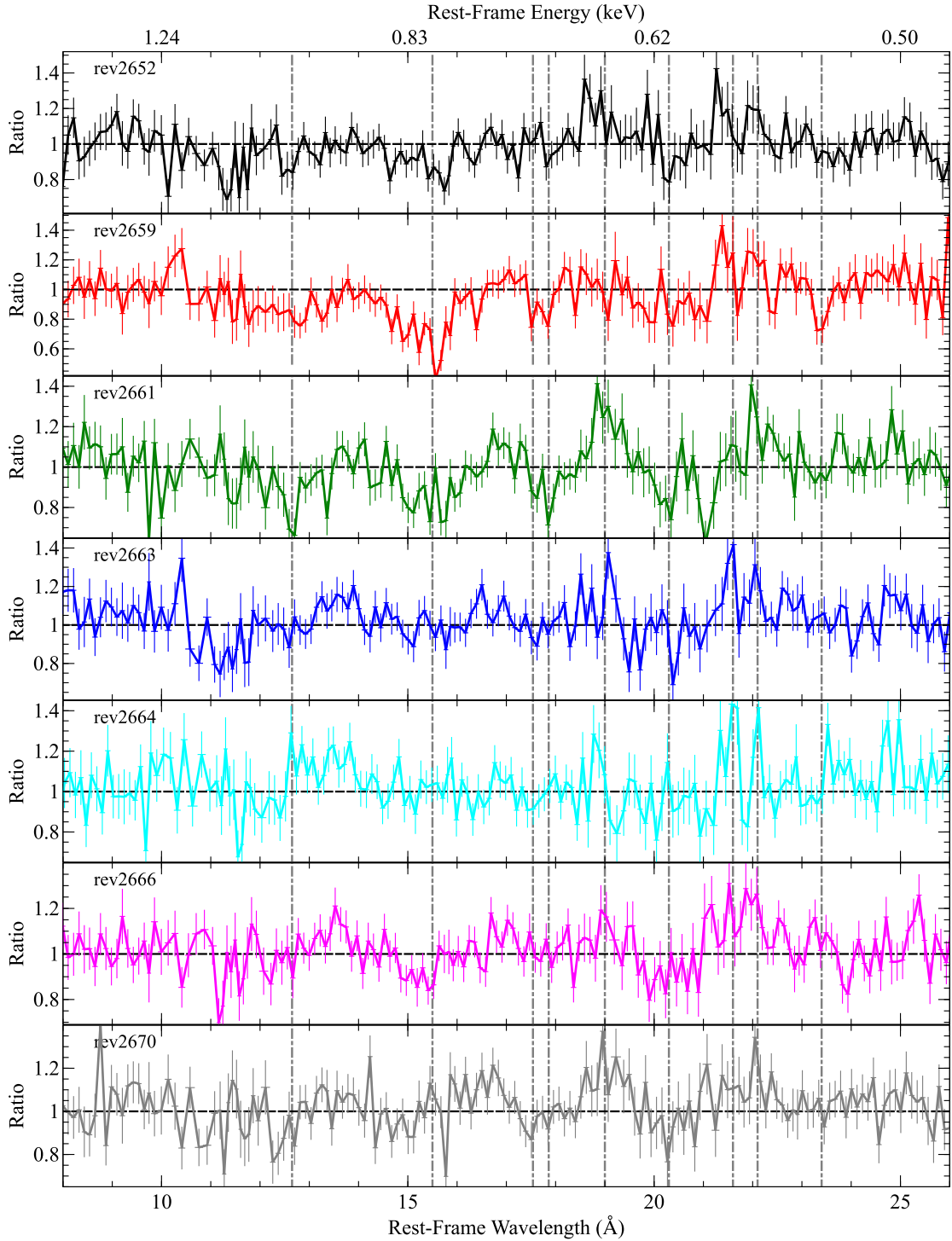}}
\end{center}
\caption{The ratio of the residuals of the seven RGS spectra to the baseline continuum model.  The vertical dot-dashed lines correspond to the centroids of:- Ne\,IX $1s\rightarrow 2p$ ($\sim12.6$\,\AA), Fe M UTA ($\sim15.5$\,\AA), O\,\textsc{vii} $1s\rightarrow 3p$ ($\sim17.6$\,\AA), O\,\textsc{viii} $1s\rightarrow2p$ 
($\sim17.9$\,\AA), O\,\textsc{viii} Ly$\alpha$ emission ($\sim$19\,\AA), O\,\textsc{vii} $1s\rightarrow2p$ 
($\sim19.2$\,\AA), O\,\textsc{vii} He$\alpha$ emission ($\sim21.6$\,\AA\ and $\sim22.1$\,\AA) and 
N\,\textsc{vii} $1s\rightarrow 2p$ ($\sim23.3$\,\AA). Of the seven spectra, \textsc{rev}\,2659 
shows the strongest absorption coinciding with the minimum in soft X-ray flux. Notice, in particular, the 
evolution of the UTA, which is at its maximum depth during \textsc{rev}\,2659, but then is not present during \textsc{rev}\,2664-2666 when the continuum flux reaches a maximum. Note also the strong variability of the O\,\textsc{viii} 
Ly$\alpha$ emission line, which is strongest in \textsc{rev}\,2661, two orbits after the decline in flux.}
\label{fig:rgs_ratio_residuals}
\end{figure*} 
 
Only two of the absorption zones are required in order to fit the sequences from the individual RGS sequences; namely the low-ionization zone 1a and the high-ionization zone 1b, both of which are outflowing at $-0.06c$ and 
which carry most of the opacity of the absorber. Note that the more tentative higher-velocity components 
(zones 2 and 3) are not required to fit the individual sequences given their lower signal-to-noise 
and so are not included in the template model. 
Likewise, the emission is also simplified, so that the O\,\textsc{vii} (forbidden and resonance) and 
O\,\textsc{viii} Ly$\alpha$ lines are fitted with simple Gaussian profiles, as these are the strongest lines that 
are detected in the individual orbit sequences. 

The continuum is modelled in the same way as per 
the mean spectral analysis, with the combination of the soft and hard power law components (both of variable normalization). 
As before, the absorption zones are applied against the hard powerlaw, while the soft powerlaw is assumed to be unattenuated. 
While a blackbody component was also initially included, its normalization dropped to zero and thus the 
continuum can simply be modeled by the combinations of the two powerlaws.
The overall form of this simple  model is:-\\

\noindent $\textsc{tbabs} \times [\textsc{xstar}_{1a} \times \textsc{xstar}_{1b} \times (\textsc{pow}_{\textsc{hard}}) + \textsc{pow}_{\textsc{soft}} + \textsc{gauss}]$,\\

\noindent where $\textsc{xstar}_{1a}$ and $\textsc{xstar}_{1b}$ represents the absorption from zones 1a and 1b respectively and \textsc{gauss} represents the Gaussian components of the O K-shell emission lines.

\begin{figure*}
\begin{center}
\rotatebox{-90}{\includegraphics[height=18cm]{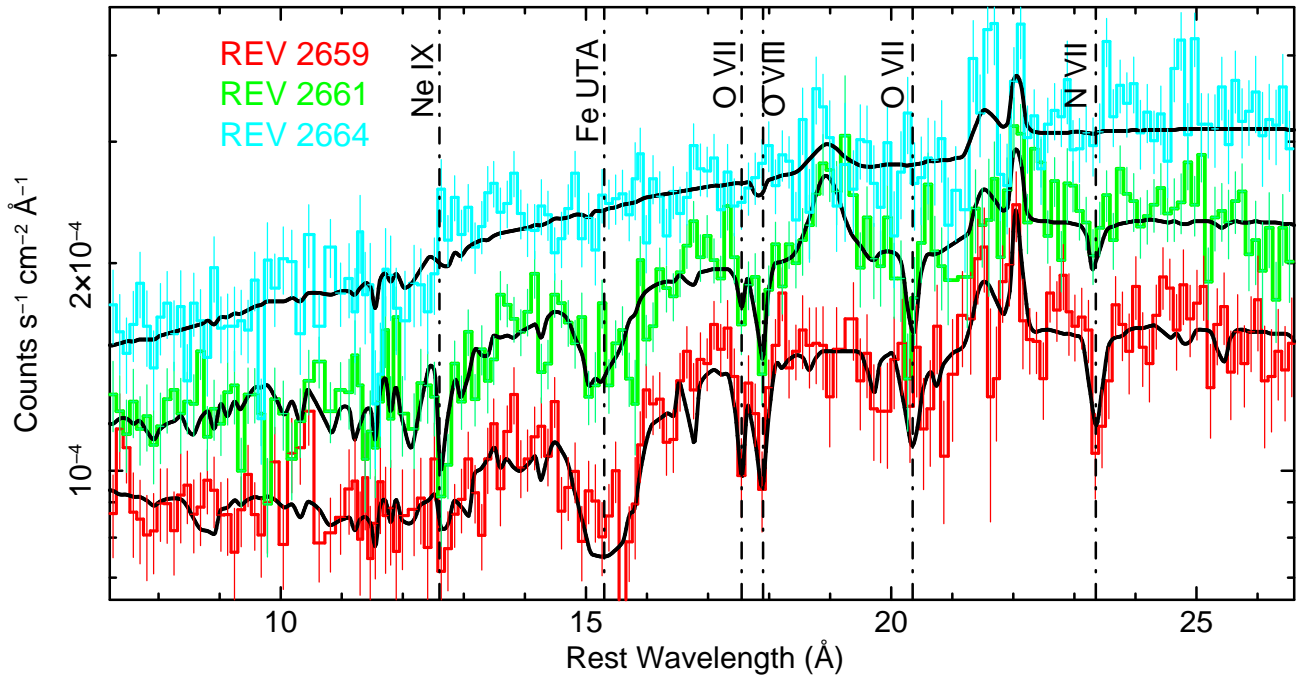}}
\end{center}
\caption{Count rate RGS spectra, showing \textsc{rev}\,2659 (red), \textsc{rev}\,2661 (green) and \textsc{rev}\,2664 (cyan) where the 
spectral changes are most apparent.
Data are shown as histograms with $1\sigma$ errors and the best-fitting \textsc{xstar} 
model to each sequence is shown as a solid line. 
The spectra are plotted following the increase in flux from \textsc{rev}\,2659 (minimum) through to \textsc{rev}\,2664 (maximum), where the soft X-ray spectrum evolves from strongly-absorbed to continuum-dominated. 
In particular, the decrease in opacity of the UTA ($\sim$15.5\,\AA), O\,\textsc{viii} Ly$\alpha$ (17.9\,\AA), O\,\textsc{vii} He$\alpha$ (20.3\,\AA) and N\,\textsc{viii} Ly$\alpha$ (23.4\,\AA) absorption features (dot--dashed lines) over time is clearly apparent.}
\label{fig:orbits}
\end{figure*}

The spectral residuals to each sequence against this baseline continuum, prior to including the absorption zones or emission lines, are shown in Figure~\ref{fig:rgs_ratio_residuals}.  The second observation (\textsc{rev}\,2659) is clearly the most absorbed of the seven spectra, which also coincides with the minimum in soft X-ray flux (e.g. see Figure~\ref{fig:xrt_pn_lc_flux}).  Conversely, the brightest observation (\textsc{rev}\,2664) also appears to be the most featureless in terms of absorption.  The most obvious spectral change among the seven spectra corresponds to the UTA, centered around $\sim$15.5\,\AA, which shows significant evolution over the course of the campaign.  In particular, it increases in strength after the first sequence, reaching maximum opacity during \textsc{rev}\,2659, before decreasing again during the subsequent observations (timescale of $\sim$days) and then becoming largely featureless when the source brightens towards a maximum flux (i.e. \textsc{rev}\,2664-2666).  To help visualise the spectral variations, we show the count rate spectra for three of the sequences (\textsc{rev}\,2659, 2661, 2664) in Figure~\ref{fig:orbits}, coinciding with the increase in continuum flux and having been fitted with the best-fitting variable column density \textsc{xstar} model described below.  This shows the spectral evolution from lowest to highest soft X-ray flux as the source changes from being absorption-dominated to continuum-dominated and largely featureless.  The clearest visualisation of this is again in the UTA, whose strength diminishes as the source brightens.  With the principal soft-band spectral changes established, we now proceed to fit the seven RGS sequences.

\subsection{Column Variability} \label{sec:column_variability}

We firstly investigated whether the inter-orbit variability could arise from changes in the column density of the absorber over the course of the seven observations.  To test this, we allowed the column density to vary between the seven sequences while tying the ionization parameter, $\xi$, and redshift (and therefore outflow velocity) of each absorber together across all seven orbits. 
We also tied the photon indices of the power laws together across observations along with the centroid energies and widths of the Gaussian emission lines. Thus the only parameters allowed to vary across the seven sequences are the column densities of the two 
absorbers, the normalizations of the two powerlaws and the flux of the strong O\,\textsc{viii} Ly$\alpha$ 
emission line (as described later).

The joint multi-orbit fit is good ($\chi_{\nu}^2=4\,592/4\,262$), with the bulk of the inter-orbit variability well accounted for by the variations in the column density. In addition, the intrinsic continuum flux increases from \textsc{rev}\,2659 (minimum) to \textsc{rev}\,2664 (maximum), 
which is reflected by the increase in the normalization of the soft powerlaw component, while the column density simultaneously decreases 
(see Table~\ref{tab:rgs_parameters} for parameters). Note there is no statistical requirement to vary the outflow velocity across the 
sequences and this was subsequently tied for all the spectra.

\begin{figure}
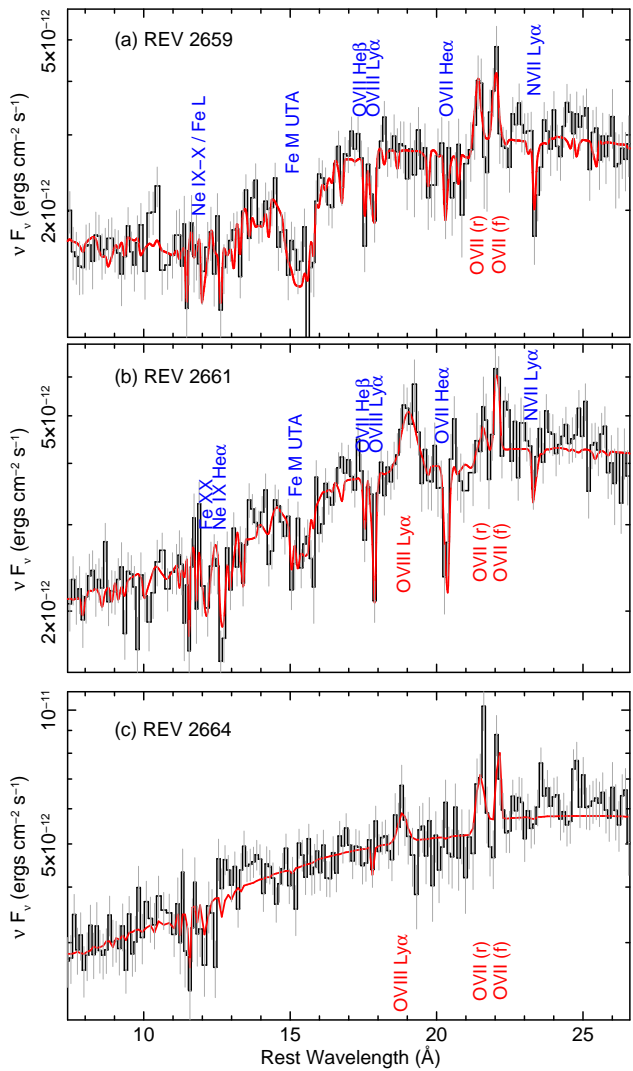

\begin{center}
\rotatebox{-90}{\includegraphics[height=8.7cm]{f9a.eps}}
\rotatebox{-90}{\includegraphics[height=8.7cm]{f9b.eps}}
\rotatebox{-90}{\includegraphics[height=8.7cm]{f9c.eps}}
\end{center}
\caption{The best-fitting two zone \textsc{xstar} model (red line), fitted to the RGS spectra of PG\,1211+143 during the 
rise in quasar flux from (a) \textsc{rev}\,2659, to (b) \textsc{rev}\,2661, through to (c) \textsc{rev}\,2664. The spectra are fitted 
allowing the column densities and continuum normalizations to vary between orbits. During \textsc{rev}\,2659, 
the low-ionization absorber (zone\,1a) is at its deepest, which can be seen in the form 
of the deep Fe M-shell UTA at $\sim$15\,\AA. In \textsc{rev}\,2661, the depth of the low-ionization UTA decreases, 
although strong, high-ionization absorption lines from O\,\textsc{vii}, O\,\textsc{viii} and Ne\,\textsc{ix} 
are still apparent from the higher-ionization zone 1b. Then in \textsc{rev}\,2664, when the quasar flux is at its maximum, 
the column density has decreased with only a trace amount of high ionization iron L-shell absorption present. 
Note the increase in flux of the broad O\,\textsc{viii} 
Ly$\alpha$ emission line at 19\,\AA\ as the quasar flux increases from \textsc{rev}\,2659 to \textsc{rev}\,2661 and 
then the subsequent decline in \textsc{rev}\,2664 at the maximum continuum flux.}
\label{fig:rev2659_rev2661}
\end{figure}

Note that the improvement in fit statistic 
for adding both absorption zones is considerable; measured the seven sequences, then $\Delta\chi^{2}=-340$ for zone 1a and $\Delta\chi^{2}=-111$ for zone 1b.
In Figure~\ref{fig:rev2659_rev2661}, we show the model fitted to the two most absorbed spectra: \textsc{rev}\,2659 and \textsc{rev}\,2661 and 
by contrast the least absorbed spectrum, \textsc{rev}\,2664.  It can be seen that the model accounts for the Fe M-shell UTA ($\sim$15.5\,\AA) along with the strongest discrete absorption lines -- e.g. from O\,\textsc{vii} and O\,\textsc{viii} Ly$\alpha$.  The variable strength of the UTA can be seen between the three orbits, which is accounted for by variations in the column density of the absorber.

\begin{deluxetable*}{l c c c c c c}
\tablecaption{Best-fitting model parameters from the multi-orbit RGS fit, fitted with either variable column density or variable covering fraction.}
\tablewidth{0pt}
\tablehead{\colhead{Obs.} & \colhead{$N^{a}_{\rm H}$ (Zone\,1a)} & \colhead{$N^{b}_{\rm H}$ (Zone\,1b)} & 
\colhead{$F_{\rm cov}^{c}$} & 
\colhead{$N^{d}$ (\textsc{POW}$_{\rm HARD}$)} & \colhead{$N^{d}$ (\textsc{POW}$_{\rm SOFT}$)} & \colhead{$F^{e}_{\rm 0.4-2 keV}$}}
\startdata
\textsc{rev}\,2652 & $4.21_{-2.23}^{+3.04}$ & $2.22_{-2.22}^{+3.80}$ & $0.17^{+0.05}_{-0.05}$ & $7.31^{+0.66}_{-0.65}$ & $10.07^{+0.69}_{-0.69}$ & $5.39^{+0.10}_{-0.09}$ \\
\textsc{rev}\,2659 & $56.8_{-16.1}^{+19.8}$ & $6.71_{-5.63}^{+1.05}$ & $0.47^{+0.05}_{-0.04}$ & $7.40^{+0.87}_{-0.78}$ & $6.61^{+0.49}_{-0.51}$ & $3.43^{+0.06}_{-0.07}$ \\
\textsc{rev}\,2661 & $11.1_{-3.52}^{+6.14}$ & $25.8_{-9.79}^{+16.0}$ & $0.30^{+0.04}_{-0.05}$ & $7.59^{+0.86}_{-0.78}$ & $9.89^{+0.67}_{-0.62}$ & $4.84^{+0.09}_{-0.09}$ \\
\textsc{rev}\,2663 & $<3.33^{\rm p}$ & $13.1_{-7.08}^{+13.3}$ & $0.14^{+0.05}_{-0.06}$ & $6.39^{+0.76}_{-0.72}$ & $10.62^{+0.66}_{-0.64}$ & $5.50^{+0.07}_{-0.08}$\\
\textsc{rev}\,2664 & $< 1.14^{\rm p}$ & $3.34_{-2.25}^{+4.87}$ & $<0.08$ & $7.53^{+0.86}_{-0.79}$ & $14.72^{+0.78}_{-0.76}$ & $6.75^{+0.11}_{-0.11}$\\
\textsc{rev}\,2666 & $5.67_{-4.00}^{+2.94}$ & $3.03_{-3.02}^{+5.20}$ & $0.11^{+0.05}_{-0.05}$ & $6.15^{+0.68}_{-0.67}$ & $15.21^{+0.71}_{-0.70}$ & $6.37^{+0.08}_{-0.08}$\\
\textsc{rev}\,2670 & $4.31_{-3.24}^{+2.26}$ & $15.2_{-7.0}^{+11.9}$ & $0.20^{+0.05}_{-0.05}$ & $6.45^{+0.75}_{-0.71}$ & $10.64^{+0.66}_{-0.65}$ & $5.05^{+0.08}_{-0.07}$\\
\enddata
\tablenotetext{a}{The column density of zone\,1a (with constant log\,$\xi = 1.8 \pm 0.1$; $v/c = -0.060 \pm 0.001$) in units of $\times 10^{20}$\,cm$^{-2}$.  $^{p}$ denotes that the column density is at the lowest value allowed by the model (i.e. $10^{19}$\,cm$^{-2}$).}
\tablenotetext{b}{The column density of zone\,1b (with constant log\,$\xi = 3.3 \pm 0.1$; $v/c = -0.062 \pm 0.001$) in units of $\times 10^{21}$\,cm$^{-2}$.}
\tablenotetext{c}{Variable covering fraction of gas, with constant column density and ionization.}
\tablenotetext{c}{The normalization of the specified power law component in units of $\times 10^{-4}$\,ph\,cm$^{-2}$\,s$^{-1}$.}
\tablenotetext{d}{The observed flux from 0.4--2\,keV in units of $\times 10^{-12}$\,erg\,cm$^{-2}$\,s$^{-1}$.} 
\label{tab:rgs_parameters}
\end{deluxetable*}

The column density variations across the seven observations are listed in Table~\ref{tab:rgs_parameters} and are plotted in Figure~\ref{fig:nh_variability} (panels a and b).  The column densities of both zones appear to vary by up to an order of magnitude across all seven observations, 
in addition to the changes in the normalizations of the powerlaw continuum components. 
The variations in the column density of the lower-ionization zone 1a appear to inversely track the flux behaviour of the source, with the column reaching a maximum value of $N_{\rm H} = 5.7^{+2.0}_{-1.6} \times 10^{21}$\,cm$^{-2}$ when the observed flux is weakest in \textsc{rev}\,2659.  
Conversely, the minimum column density occurs during the brighter sequences -- e.g. \textsc{rev}\,2664, with only an upper limit of $N_{\rm H} < 1.1 \times 10^{20}$\,cm$^{-2}$.  The high column of \textsc{rev}\,2659 corresponds to an increase in the strength of the UTA at around $\sim$15--16\,\AA, whereafter the column density decreases over the subsequent observations as the source flux increases.  The variability of the absorber is observed to be rapid with the column density of zone\,1a decreasing by a factor of $\sim$5 from \textsc{rev}\,2659 to \textsc{rev}\,2661 - a period of just $\sim$4\,days. The changes in absorber opacity vs flux can clearly be seen comparing the respective panels of
Figure~\ref{fig:nh_variability} (panels a and b vs e).

Similar behaviour is observed in the higher-ionization zone 1b, with an increase in the column density required to match an increased strength in high-ionization absorption lines when the source flux is lower.  However, in the case of zone\,1b, the column density actually peaks during \textsc{rev}\,2661 ($N_{\rm H} = 2.6^{+1.6}_{-1.0} \times 10^{22}$\,cm$^{-2}$) and thus is delayed in its response to the variable source continuum with respect to the lower-ionization zone 1a, where the column density peaks a couple of orbits earlier.  This behaviour is also apparent in Figure~\ref{fig:rev2659_rev2661}, where the \textsc{rev 2661} spectrum shows strong highly ionized absorption lines due to O\,\textsc{vii}, O\,\textsc{viii} and Ne\,\textsc{ix}, whereas in \textsc{rev 2659} the low ionization Fe UTA is at its strongest. This might imply two separate obscuration events occurring along the line of sight which are spatially independent, or a co-spatial two phase medium whereby the high ionization absorber reacts differently to the lower ionization gas.  Nevertheless, in both cases, the column densities further decrease when the observed source continuum flux reaches its maximum value around \textsc{rev}\,2664-2666 -- i.e. imprinting minimal features onto what becomes a largely featureless soft X-ray spectrum.

\subsection{Covering Fraction Variations} \label{sec:covering}

Given the rapid variability of the column density of the absorbing gas in PG\,1211+143, it seems plausible that the increase in opacity around \textsc{rev}\,2659 could arise from the passage of patchy absorbing clouds or clumps of gas into our line of sight, as 
part of an inhomogeneous wind. 
This can be tested by modeling the opacity changes by a variable partially covering absorber, 
such that part of the continuum leaks through a patchy medium and only part is attenuated.
For this purpose, the model construction is altered such that:-\\

\noindent $\textsc{tbabs} \times [c_1 \times \textsc{xstar}_{1a} \times \textsc{xstar}_{1b} \times (\textsc{pow}_{\textsc{hard}} + \textsc{pow}_{\textsc{soft}}) + c_2 \times (\textsc{pow}_{\textsc{hard}} + \textsc{pow}_{\textsc{soft}}) + \textsc{gauss}]$,\\ 

where the covering fraction is then $f_{\rm cov}=c_1/(c_1 + c_2)$ and the other components are as before. 
The column densities are tied across the seven sequences (zone\,1a, $N_{\rm H}=5.4^{+1.9}_{-1.2}\times10^{21}$\,cm$^{-2}$; zone\,1b, $N_{\rm H}=3.7^{+3.1}_{-1.2}\times10^{22}$\,cm$^{-2}$), 
while the ionizations and velocities of the two zones are also tied with values found to be consistent from before. Note the relative powerlaw normalizations vary between sequences and thus the 
flux is allowed to be intrinsically variable.
The resulting covering fraction variations are plotted in Figure~\ref{fig:nh_variability}, panel (c) and listed in Table~\ref{tab:rgs_parameters}; these vary from the highest covering during \textsc{rev}\,2659 ($f_{\rm cov}=0.47^{+0.05}_{-0.04}$) and lowest during the featureless \textsc{rev}\,2664 ($f_{\rm cov}<0.08$) spectrum.

The covering fraction is found to be inversely proportional to the quasar flux (panel c vs panel e)  
and thus the quasar is most absorbed when it is faintest. Note that the absorber variations are similar 
to those found for the variable column density case, while the fit statistic is virtually identical with 
$\chi_{\nu}^{2}=4\,589/4\,259$.
This is not surprising, as in practice, variations in the covering factor (for a given cloud column density) will be indistinguishable from the equivalent variations 
in column density (for a given covering) of the absorber.
The line of sight column / covering variations are discussed further in Section~\ref{sec:location_properties_gas}.

\subsection{Continuum Variations Only} \label{sec:continuum_variability}

As a further consistency check, we also attempt to see whether the spectral changes between the sequences can be accounted 
for purely by continuum variability of the soft vs hard powerlaw components. In the original model construction, only the direct hard power-law 
is absorbed, while the soft power-law is unattenuated; thus it may be possible to mimic the opacity changes by an increase in the 
soft component, relative to a less variable hard powerlaw component. For instance, in the lowest flux spectrum (\textsc{rev}\,2659) 
the soft flux is low and the spectrum appears more absorbed, while in the brightest sequences (e.g. \textsc{rev}\,2664), the flux of the 
soft component strongly increases with respect to the hard power-law. Indeed such a behaviour was noted in the {\it Chandra} 
observations of the Seyfert\,1 galaxy, NGC\,3783 \citep{Netzer03}, where although no changes in the warm absorber were observed, a featureless 
soft component was found to vary between the observations.

To test this, we tied the column densities of the two absorbers across all seven sequences, allowing only the normalizations of the 
respective soft vs hard powerlaw components to vary between the seven orbits. However the result was a substantially worse 
fit compared to the variable column density case (where both the continuum normalizations and the column densities were allowed to vary across 
all the datasets). Indeed, the fit statistic worsens to $\chi^{2}_{\nu}=4\,742.9/4\,274$, resulting in a $\Delta\chi^2=+151.5$ for 
12 extra degrees of freedom. Such an increase in the fit statistic is strongly ruled out, with a null hypothesis probability of 
$1.5\times10^{-23}$ and thus continuum variations alone cannot explain the spectral variability across the seven observations of 
PG\,1211+143. In particular, this scenario fails to reproduce the strong changes around the Fe M-shell UTA, 
which is observed from \textsc{rev}\,2659 to \textsc{rev}\,2664.  

\subsection{Ionization Variability} \label{sec:ionization_variability}

While pure continuum variability is ruled out, 
we tested whether the inter-orbit variability could be due to variability of the absorber ionization in response to the variable continuum, instead of being driven by large column changes.
Thus, the column density of each absorber was tied to the same 
value across all seven orbits for both zones 1a and 1b, while the ionization was allowed to vary instead. 
However the resulting fit statistic was found to be worse by $\Delta\chi^2=40$ compared to the above 
variable column density scenario.  
In particular, variations in ionization alone cannot reproduce the increase in strength of the UTA arising
from the low-ionization (zone\,1a) absorber, which instead requires a significant increase in column, especially during the low flux \textsc{rev}\,2659 sequence.

\begin{figure*}
\begin{center}
\rotatebox{-90}{\includegraphics[height=16cm]{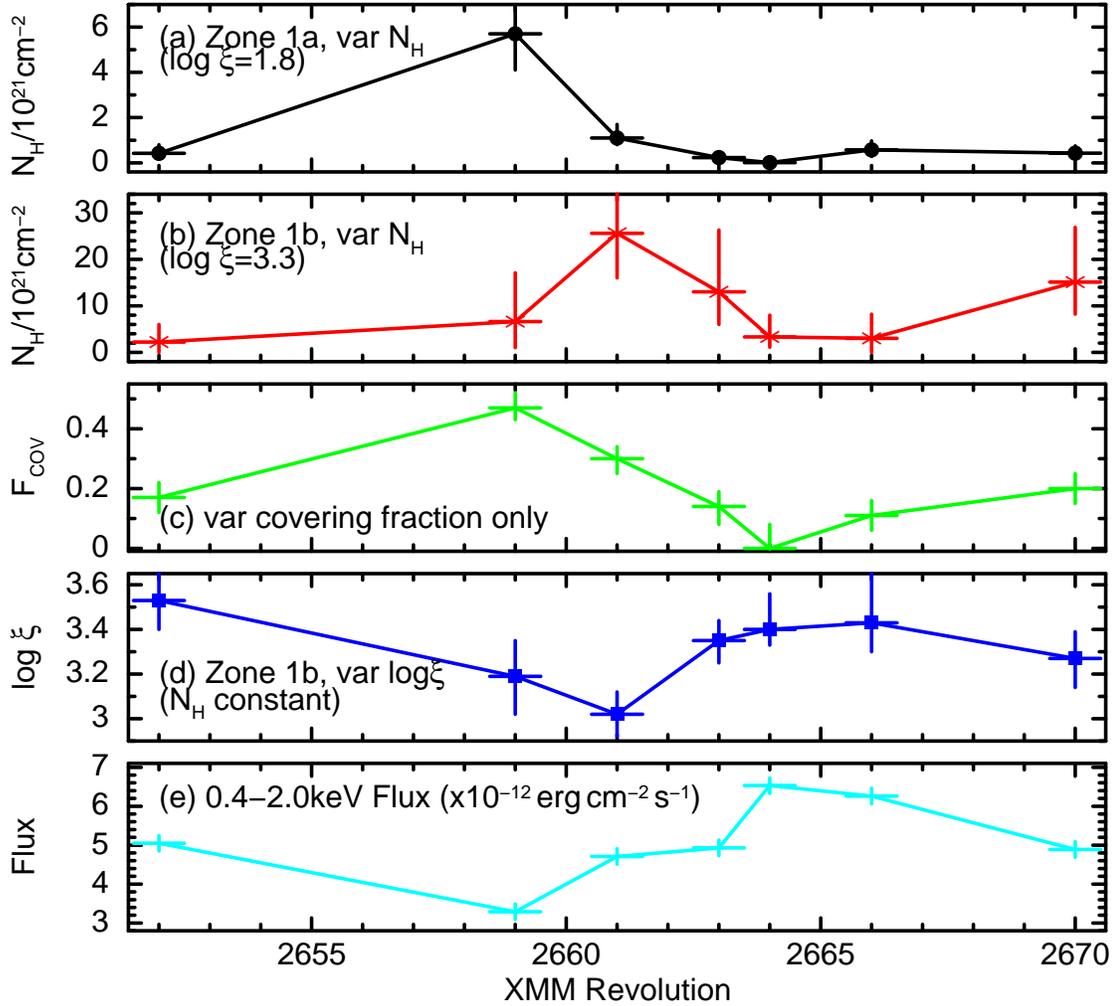}}
\end{center}
\caption{The variability of the two high-opacity absorbers (zones 1a and 1b) across the seven {\it XMM-Newton} RGS observations (see Table~\ref{tab:rgs_parameters} and Table~\ref{tab:rgs_ionization} for details of the absorber variability). 
Panel (a) shows the variations in column density for the low-ionization zone 1a, which 
reaches its peak in \textsc{rev}\,2659 during the low-flux episode, $\sim$15 days into the campaign. 
The column then rapidly declines in subsequent orbits as the quasar flux increases. 
Panel (b) shows a similar behaviour for the high-ionization zone 1b, except here, the column density reaches a maximum two orbits later during \textsc{rev}\,2661. Panel (c) shows the absorber variations instead expressed in terms of a variable covering fraction, but assuming a constant column and ionization across the seven sequences. Note the covering anti-correlates with 
the flux in panel (e).
Panel (d) shows the zone 1b variations, 
where here, the absorber variations have instead been fitted with a variable ionization parameter, 
but a constant column density. This may indicate that the high-ionization absorber reacts to changes in the 
soft X-ray continuum, reaching a minimum ionization shortly after the decline in continuum quasar flux shown in panel (e). }
\label{fig:nh_variability}
\end{figure*}

Thus the variability of the low-ionization absorber appears primarily 
driven by column density changes, although the variability of the high-ionization absorber may still be accounted for by variable ionization. To test this hybrid scenario, we allowed the 
ionization parameter to vary for the high-ionization (zone\,1b) absorber (with a constant column density), 
while, conversely, the column density of the low-ionization (zone\,1a) absorber was allowed to vary as before (and assuming a constant ionization parameter). This gave a better fit, with $\chi_{\nu}^2=4\,590/4\,262$, statistically equivalent to the purely variable column density case from Section~\ref{sec:column_variability}. 
Here, the column density of the low-ionization absorber varies as above, but the ionization parameter of the highly-ionized absorber appears to track the soft X-ray continuum variability, which can be seen from comparing panels (d) and (e) in Figure~\ref{fig:nh_variability}.

The resulting ionization changes for zone\,1b 
are shown in Figure~\ref{fig:nh_variability} (panel d) and listed in Table~\ref{tab:rgs_ionization}, with the ionization parameter at its lowest  
value of $\log\xi=3.02\pm0.10$ during \textsc{rev}\,2661, when the high-ionization absorption lines appear strongest in the RGS spectrum, as is apparent in Figure~\ref{fig:rev2659_rev2661}. 
Subsequently, the absorber ionization increases when the 
continuum flux reaches its maximum value during \textsc{rev}\,2664 and \textsc{rev}\,2666, with $\log\xi=3.40^{+0.16}_{-0.07}$ in the former sequence. 
The increase in ionization parameter leads to the high-ionization absorber becoming more transparent 
and, hence, the higher-flux sequences are relatively featureless.
Note that the magnitude of the changes in ionization for zone\,1b are roughly in proportion to the changes in continuum flux, with the latter increasing by a factor of 2 from \textsc{rev}\,2659 to \textsc{rev}\,2664. There may also be some indication of a delay in the ionization in response to the continuum variability, 
with the minimum in ionization occurring during \textsc{rev}\,2661 when the continuum flux has already started to increase, while the minimum in flux occurs 
$\sim$2 {\it XMM-Newton} orbits earlier during \textsc{rev}\,2659. This can be seen from comparing panels (d) and (e) in Figure~\ref{fig:nh_variability}.

In summary, the variability of the low-ionization absorber is most likely to be accounted for by changes 
in column density (or equivalently covering fraction),
 with the quasar most absorbed during the low flux \textsc{rev}\,2659 observation. 
On the other hand, the variability of the high-ionization absorber can be equally well explained 
by either an increase in column density, centered on \textsc{rev}\,2661, or by changes in the absorber ionization in response to the variable soft X-ray continuum.

Finally, we note that, in emission, the broad O\,\textsc{viii} Ly$\alpha$ line appears highly variable, 
with the line fluxes and equivalent widths (EWs) listed for each sequence in Table~\ref{tab:rgs_ionization}. 
Note that no significant variability is observed in the O\,\textsc{vii} emission by comparison, 
although the larger errors preclude a direct comparison.
In particular, the O\,\textsc{viii} emission line flux drops in the low-flux \textsc{rev}\,2659 
spectrum and then increases two orbits later (in \textsc{rev}\,2661) as the continuum flux rises, which 
is visible in Figure~\ref{fig:rev2659_rev2661}.
Quantitatively, the line flux (and EW) varies from $<1.2\times10^{-5}$\,photons\,cm$^{-2}$\,s$^{-1}$ (${\rm EW}<2.0$\,eV) to $4.8\pm1.1\times10^{-5}$\,photons\,cm$^{-2}$\,s$^{-1}$ (${\rm EW}=6.0\pm1.4$\,eV) between these two orbits.
However, in the subsequent orbits (e.g. \textsc{rev}\,2664), 
as the continuum continues to rise, the line flux (and EW) appear to decrease. 
Thus, the emission line does not directly respond in proportion to the continuum, as might be 
expected in a simple reverberation scenario, which 
would lead to a constant equivalent width of the emission line.
Alternatively, as is discussed in Section~\ref{sec:location_properties_gas}, the increase in line flux in \textsc{rev}\,2661 may be explained in terms of a delay due to recombination following the earlier drop in continuum flux centered on \textsc{rev}\,2659. In the subsequent orbits, the decline in the emission line strength may then be due to an increase in the ionization, in response to the rise in the continuum flux.

\begin{deluxetable}{l c c c c}
\tablecaption{Best-fitting model parameters for the high-ionization zone 1b absorber, when the ionization parameter is allowed to vary between the seven observations.  Note here that the column density has been tied across seven observations, with a best-fitting value of $N_{\rm H}=7.0^{+3.0}_{-2.0}\times10^{21}$\,cm$^{-2}$.}
\tablewidth{0pt}
\tablehead{\colhead{Obs.} & \colhead{log\,$\xi$} & \colhead{O\,\textsc{viii} Flux$^{a}$} & \colhead{O\,\textsc{viii} EW$^{b}$} & \colhead{$F^{c}_{\rm 0.4-2 keV}$}}
\startdata
\textsc{rev}\,2652 & $3.53^{+0.17}_{-0.13}$ & $3.3^{+1.2}_{-1.2}$ & $4.0^{+1.5}_{-1.5}$ & $5.39^{+0.10}_{-0.09}$ \\
\textsc{rev}\,2659 & $3.19^{+0.16}_{-0.17}$ & $<1.2$ & $<2.0$ & $3.43^{+0.06}_{-0.07}$ \\
\textsc{rev}\,2661 & $3.02^{+0.10}_{-0.10}$ & $4.8^{+1.1}_{-1.1}$ & $6.0^{+1.4}_{-1.4}$ & $4.84^{+0.09}_{-0.09}$ \\
\textsc{rev}\,2663 & $3.35^{+0.09}_{-0.10}$ & $1.8^{+1.1}_{-1.1}$ & $2.2^{+1.3}_{-1.3}$ & $5.50^{+0.07}_{-0.08}$\\
\textsc{rev}\,2664 & $3.40^{+0.16}_{-0.07}$ & $1.4^{+1.5}_{-1.4}$ & $1.3^{+1.4}_{-1.3}$ & $6.75^{+0.11}_{-0.11}$\\
\textsc{rev}\,2666 & $3.43^{+0.22}_{-0.13}$ & $1.7^{+1.2}_{-1.2}$ & $1.6^{+1.1}_{-1.1}$ & $6.37^{+0.08}_{-0.08}$\\
\textsc{rev}\,2670 & $3.27^{+0.12}_{-0.13}$ & $2.8^{+1.1}_{-1.1}$ & $3.3^{+1.3}_{-1.3}$ & $5.05^{+0.08}_{-0.07}$\\
\enddata
\tablenotetext{a}{The flux of the high-ionization O\,\textsc{viii} Ly$\alpha$ emission line, which appears to vary between orbits, in units of $\times 10^{-5}$\,ph\,cm$^{-2}$\,s$^{-1}$.  Note that the highest O\,\textsc{viii} flux coincides when the ionization of the absorber is lowest, during rev2661.}
\tablenotetext{b}{The equivalent width of the O\,\textsc{viii} Ly$\alpha$ emission line in units of eV.}
\tablenotetext{c}{The observed 0.4--2\,keV flux in units of $\times 10^{-12}$\,erg\,cm$^{-2}$\,s$^{-1}$.}
\label{tab:rgs_ionization}
\end{deluxetable}


\section{Discussion} \label{sec:discussion}

\subsection{Comparison to Other Fast, Soft X-ray Outflows}

We have carried out a detailed analysis of the deep 2014 {\it XMM-Newton} RGS 
observations of the ultra fast outflow in PG\,1211+143, finding a number of blue-shifted absorption lines with a common outflow velocity of $\sim-0.06c$. 
We note that recently, \citet{Kriss17} have discovered a UV component of the fast outflow of PG\,1211+143 from the detection of a Ly$\alpha$ trough, where the strongest component has 
an outflow velocity of -$17420\pm15$\,km\,s$^{-1}$. 
This compares to the main zone\,1a and 1b components 
of the wind found in this RGS analysis (Section 3.2), which have respective outflow velocities of 
$-0.062\pm0.001c$ (or $-18600\pm300$\,km\,s$^{-1}$) and $-0.059\pm0.002c$ (or $-17700\pm600$\,km\,s$^{-1}$). In particular the UV absorption line  system measured by \citet{Kriss17} is consistent within the errors with the velocity of the zone\,1b absorber measured here. 
Furthermore an analysis of the simultaneous Chandra HETG observations of PG\,1211+143 by \citet{Danehkar17}, taken one year later compared to the 2014 {\it XMM-Newton} observations, 
also confirms the detection of the $-0.06c$ component of the fast soft X-ray outflow, through the 
presence of absorption lines detected in the Ne, Mg and Si bands.

While the most ultra fast outflows have been detected in the iron K band through 
high ionization lines of Fe\,\textsc{xxv} or Fe\,\textsc{xxvi} \citep{Tombesi10, Gofford13}, 
PG\,1211+143 is one among a small number AGN, with sufficient exposure with X-ray gratings, where components of an ultra fast outflow have also been detected in the soft X-ray band \citep{Pounds14,Pounds16b}. Now, 
both the current {\it XMM-Newton} analysis and the {\it Chandra} analysis of \citet{Danehkar17}, 
confirm the presence of a fast soft X-ray wind in PG\,1211+143, with consistent velocities. 
Examples of fast soft X-ray outflows in other AGN include the luminous nearby quasar PDS\,456 \citep{ReevesOBrienWard03, Reeves16}, the NLS1s IRAS\,17020+4544 \citep{Longinotti15}, IRAS\,13224+3809 \citep{Parker17, Pinto17a} and Ark 564 \citep{Gupta13}, as well as the Seyfert 1 galaxy Mrk\,590 \citep{GuptaMathurKrongold15}. 

Of these AGN, the NLS1 IRAS\,17020+4544 \citep{Longinotti15}
may bare the closest resemblance to PG\,1211+143, in terms of the complexity of the soft X-ray outflow. PG\,1211+143 shows clear contributions from both low and high ionization zones of the fast soft X-ray absorber, covering two orders of magnitude in ionization, where the imprint of the outflow ranges from the low ionization Fe M-shell UTA, to H-like absorption from N, O and Ne. Likewise, the outflow in IRAS\,17020+4544 also exhibits a wide range in ionization, especially in the O K-shell band, with significant low ionization absorption occurring from the inner shell lines of O\,\textsc{iii-vi}, with 
higher ionization absorption from O\,\textsc{vii-viii} also being present.
\citet{Longinotti15} discuss the fast, but low ionization, absorption as arising from either a clumpy wind, or from the immediate post shock gas following interaction of a faster (more ionized) wind component with circumnuclear gas (see \citealt{King10}). 
The observed outflow velocity is also similar to PG\,1211+143, with possible multiple velocity components, ranging from $-23600$\,km\,s$^{-1}$ to $-27200$\,km\,s$^{-1}$ for the three most significant zones. 

In contrast to IRAS\,17020+4544, where a spread of velocities may be observed, in PG\,1211+143, 
most of the absorption from both the low and high ionization gas originates from lines found with a common mean velocity of $-0.062\pm0.001c$ (or $-18600\pm300$\,km\,s$^{-1}$), and subsequently confirmed from the
 \textsc{xstar} analysis in Section 3.2 (see Table~4, absorber zones\,1a and 1b). 
There is some indication in the 2014 data of a slightly higher velocity component, with $v=-0.077\pm0.001c$ (or $-23100\pm300$\,km\,s$^{-1}$) for the zone 2 absorber. In the spectrum, this is seen in the form of a second absorption line component observed blue-wards of the O\,\textsc{vii} He-$\alpha$ and O\,\textsc{viii} Ly$\alpha$ line profiles (Figure~\ref{fig:mean_xstar}). Although weak, this latter component consistent with the main velocity component measured in the 2001 {\it XMM-Newton} observation, 
of $-23170\pm300$\,km\,s$^{-1}$ \citep{Pounds14}.
This is formally lower than the systemic velocity of PG\,1211+143 (of $-24300$\,km\,s$^{-1}$) and so probably does not originate from Galactic absorption. 


As we have seen in Section 4, the seven 2014 {\it XMM-Newton} sequences do show pronounced variability in the absorber opacity, in addition 
to an intrinsically variable continuum. 
This may be interpreted in terms of variations in either column density, covering fraction or ionization changes, 
occurring on timescales of days. Indeed PG\,1211+143 appears to follow the same 
trend as seen towards the soft X-ray outflows in both PDS\,456 and IRAS\,13224+3809, where the absorber opacity diminishes with increasing luminosity. In IRAS\,13224+3809, \citet{Pinto17a} have recently shown that the absorption lines from the fast outflow seen in both the soft X-ray and Fe K bands decreases in strength with increasing continuum flux; this was attributed to the increasing ionization state of the wind, with the absorber essentially becoming transparent at high fluxes. Likewise, in PG\,1211+143, the 
soft X-ray spectrum becomes featureless in the highest flux sequences, such as 
in \textsc{rev 2664} and \textsc{rev 2666}. 
In PDS\,456, \citet{Reeves16} also found that the broad absorption profiles seen in the RGS data were generally stronger when the quasar was 
more absorbed and in a lower flux state, with the soft X-ray 
absorber changes attributed to a variable column density or 
covering fraction. The soft X-ray phase of the wind in PDS\,456 was interpreted as arising from denser clumps of material located within the fast wind. 
Furthermore in the iron K band, the strength of the ultra fast outflow in PDS\,456 appears to be anti-correlated with the continuum, as seen from a Principal Components Analysis of the archival {\it XMM-Newton} observations \citep{Parker17b}. Similar to IRAS\,13224+3809, this finding may be interpreted in terms of the wind ionization reacting to the continuum flux.

In PG\,1211+143, the new {\it XMM-Newton} observations appear to have resolved an absorption event, centered on \textsc{rev 2659}, the most absorbed sequence of the long look campaign. Next, we investigate
the constraints that the absorber variability and timescales provide on the properties of the soft X-ray absorber in PG\,1211+143. 

\subsection{Location and Properties of the Outflowing Gas} \label{sec:location_properties_gas}

The variability of the soft X-ray absorber can be used to deduce the properties and location of the 
outflowing gas in PG\,1211+143. The rapid variability of the column density (or covering fraction) for the low-ionization 
(zone\,1a) absorber suggests that this phase of the outflow is highly inhomogeneous. 
We can attribute the increase in column density around \textsc{rev}\,2659 to the passage of an 
absorbing cloud or filament across our line of sight. 
In that case, the timescale of the absorption event can give an estimate of the size-scale ($\Delta R$) of the absorber, as $\Delta R = v_{\rm t} \Delta t$, where $v_{\rm t}$ 
is the transverse velocity of the absorber across our line of sight. While the transverse velocity is 
not known (but is likely to be smaller than the outflow velocity along the direct line of sight), as a first approximation this can be set equal to the Keplerian velocity at a radius $R$, 
where $v_{\rm t} = (GM/R)^{1/2}$. Thus:-
\begin{equation}
\Delta R = \left(\frac{GM}{R}\right)^{1/2}\Delta t.
\end{equation}
\noindent The radius, $R$, is related to the ionization parameter and electron density of the absorber 
as $R^{2}=L/n_{\rm e}\xi$, where $L$ is the 1--1000 Rydberg ionizing luminosity. 
The electron density is also related to the absorber size-scale 
as $n_{\rm e} = \Delta N_{\rm H} / \Delta R$, where $\Delta N_{\rm H}$ is the change in column density 
measured from the observations.

Combining these expressions then gives, for the radial location of the absorber, $R$:-
\begin{equation}
R^{5/2} = (GM)^{1/2} \frac{L \Delta t}{\Delta N_{\rm H} \xi}. \label{eq:radial_location}
\end{equation}
\noindent The change in the column density of the low-ionization absorber is 
$\Delta N_{\rm H}=5\times10^{21}$\,cm$^{-2}$, occurring over a timescale of 
approximately 2 {\it XMM-Newton} orbits or $\Delta t = 300$\,ks; i.e. as measured from the decline of the column density from \textsc{rev}\,2659 to \textsc{rev}\,2661 (see Figure~\ref{fig:nh_variability}; panel a).
The ionizing luminosity of PG\,1211+143 from 
the measured optical to X-ray SED is $L=4\times10^{45}$\,erg\,s$^{-1}$ \citep{Lobban16a}, 
while the ionization of the gas is $\log\xi=1.8$ (Table~\ref{tab:rgs_parameters}). 
We assume a black hole mass of $M=10^{8} {\rm M}_{\odot}$, which is within the range of the reverberation mass estimates for PG\,1211+143 of $0.4-1.5\times10^{8} {\rm M}_{\odot}$ \citep{Kaspi00, Peterson04}. Thus, while there is some uncertainty in the exact black hole mass of PG\,1211+143, the above radius 
is only a weak function of the mass (i.e. $R\propto M^{1/5}$). 

For the above values, the estimated radial distance of the low-ionization (zone\,1a) absorber is 
$R=7\times10^{17}$\,cm or $\sim$10$^{18}$\,cm. From this, the corresponding density is $n_{\rm e}=L/\xi R^2 \sim 10^{8}$\,cm$^{-3}$ and the absorber size-scale is $\Delta R = \Delta N_{\rm H} / n_{\rm e} \sim 5\times10^{13}$\,cm, while the transverse velocity of
absorber clouds is then $v_{\rm t}=1\,500$\,km\,s$^{-1}$. Thus, in 
order to reproduce the rapid increase in column of the low-ionization 
absorber, centered on \textsc{rev}\,2659, the absorbing clouds have to be compact (a few gravitational radii in extent), relatively 
dense ($n_{\rm e}\sim10^{8}$\,cm$^{-3}$) and highly inhomogeneous (with $\Delta R/R\sim10^{-4}$).

The same argument can also be used to estimate the size-scale and location of the high-ionization (zone\,1b) absorber, from the column density variations presented earlier in Section~\ref{sec:column_variability}. 
In this case, $\Delta N_{\rm H}=2\times10^{22}$\,cm$^{-2}$ and $\Delta t=500$\,ks from the decline of zone\,1b of the absorber from \textsc{rev}\,2661 to \textsc{rev}\,2664 (three {\it XMM-Newton} orbits), while the ionization is 
$\log\xi=3.3$. Hence, from equation~\ref{eq:radial_location}, $R=7\times10^{16}\sim10^{17}$\,cm and, subsequently, 
$n_{\rm e}\sim3\times10^{8}$\,cm$^{-3}$, $\Delta R\sim10^{14}$\,cm, and the transverse velocity is 
then $v_{\rm t}=4\,000$\,km\,s$^{-1}$. Thus, the absorber location is an order of magnitude closer in 
to the black hole, which may be expected given the higher ionization of this zone and similar density to the above. Note the derived transverse velocity is then $v_{\rm t}=4\,000$\,km\,s$^{-1}$, which is not inconsistent with the measured velocity widths of the soft X-ray emission lines in the PG\,1211+143 spectrum.


\subsection{A Two-Phase Outflow?} \label{sec:two-phase_outflow}

The above argument suggests that the low- and high-ionization absorbers could arise from gas at two different radial locations, with the higher-ionization gas closer in.  
However, if the two absorber zones were physically unconnected, 
it might seen coincidental that the increase in absorption is seen from both zones during this campaign unless they both move into the line of sight at around the same time.
Furthermore, given the identical outflow 
velocities (consistent with $-0.06c$ within errors) of these two absorption zones, it may instead be more plausible that the absorbers exist co-spatially within a two phase medium. 

To test this, we consider a scenario whereby the ionization of the more highly-ionized medium 
responds directly to the continuum and where any delay in the response of the 
absorber following a decrease in the continuum level might arise as a result of the recombination 
timescale of the gas. 
As per Section~\ref{sec:ionization_variability}, in this case we assume any column density variations are negligible, which may be reasonable if the high-ionization gas is less dense and more spatially extended than for the lower-ionization clumps.
The density (and thus the gas location) can then be estimated, as it is inversely proportional to the recombination time.

From the analysis in Section~\ref{sec:ionization_variability}, we found that the change in the ionization parameter of absorption zone 1b 
may be slightly delayed with respect to the continuum. Here, the minimum in the absorber ionization parameter during \textsc{rev}\,2661 (Figure~\ref{fig:nh_variability}; panel d) occurs when the continuum was already increasing 
(panel e), approximately two {\it XMM-Newton} orbits after the minimum in the soft X-ray flux during \textsc{rev}\,2659.
This may be interpreted as a delay due to recombination, with the decrease in ionization following after a decrease in ionizing flux.
A similar behaviour is also seen for the O\,\textsc{viii} Ly$\alpha$ emission line, which reaches a maximum intensity in \textsc{rev}\,2661 (Table~\ref{tab:rgs_ionization}), following the earlier minimum in the continuum flux, perhaps arising as a result of a decrease in ionization.

The recombination time can be approximated as $t_{\rm rec} \sim (\alpha_{\rm r} n_{\rm e})^{-1}$, 
where $\alpha_{\rm r}$ is the recombination coefficient for ions recombining into O\,\textsc{viii}. 
We note the above is only a simplistic approximation for the recombination time and the resulting densities 
should only be treated as an order of magnitude estimate.
From the observations, we estimate $t_{\rm rec}=300$\,ks (i.e. two {\it XMM-Newton} orbits) as a realistic 
recombination time for the highly-ionized gas. 
On the other hand, $\alpha_{\rm r}$ is dependent on the temperature of the gas. 
This may be estimated from the observed width of any RRC features in the RGS spectrum, with the best constraints arising from the O\,\textsc{vii} RRC (see Table~\ref{tab:emission-lines}), with $kT\sim18$\,eV. Thus, for $T=2\times10^{5}$\,K, 
then $\alpha_{\rm r}=7.5\times10^{-12}$\,cm$^{3}$\,s$^{-1}$ \citep{NaharPradhan03} and, subsequently, 
$n_{\rm e}=5\times10^{5}$\,cm$^{-3}$. Note that the O\,\textsc{viii} RRC 
from the higher-ionization gas (see also \citealt{PoundsReeves09}) is not so apparent, 
which may be the case if the temperature of the more highly ionized gas is greater (and thus the RRC is too broad to detect), or it may be blended into the broad Ne\,\textsc{ix} emission line. 
In the former case, for $T=10^{6}$\,K, then $\alpha_{\rm r}=2.1\times10^{-12}$\,cm$^{3}$\,s$^{-1}$ and $n_{\rm e}=1.4\times10^{6}$\,cm$^{-3}$.

A plausible estimate for the density of the high-ionization gas is $n_{\rm e}\sim10^{6}$\,cm$^{-3}$. Thus for a mean ionization parameter of $\log\xi=3.3$, then the typical radial 
distance is $R\sim10^{18}$\,cm. This is in agreement with the radial estimate derived for the compact 
low-ionization 
absorbing clouds. Thus, the data are consistent with a two-phase outflowing medium, whereby the high-density, but low-ionization, clouds can co-exist within a lower-density, but higher-ionization, medium. In this scenario, both of these absorbers may form part of the same wind streamline outflowing at $-0.06c$.  

\subsection{Mass Outflow Rate and Energetics}

With an estimate of the likely outflow radius and the overall properties of the gas, it is then possible to calculate the possible
contribution of the soft X-ray outflow towards the mass outflow rate. This can be expressed in the following form (e.g. see \citealt{Blustin05}):-
\begin{equation}
\dot{M}_{\rm out}=4\pi f_{\rm cov} \mu m_{\rm p} v_{\rm out} f_{\rm v} n_{\rm e} R^{2}. \label{eq:mass_outflow_rate}
\end{equation}
Here, $f_{\rm cov}$ corresponds to the global covering factor of the outflow (i.e. as a fraction of 
$4\pi$\,steradians solid angle), $\mu m_{\rm p}$ is the average atomic mass of the gas, while $f_{\rm v}$ is the 
volume filling factor along the flow. For a more homogeneous outflow (where $f_{\rm v}\sim1$), then the above expression can be integrated with respect to radius to instead obtain the mass outflow rate in terms of the column density (e.g. \citealt{Nardini15}):-
\begin{equation}
\dot{M}_{\rm out}=4\pi f_{\rm cov} \mu m_{\rm p} v_{\rm out} N_{\rm H} R, \label{eq:mass_outflow_rate_homogeneous}
\end{equation}
where, here, $R$ then corresponds to the inner radial distance to the wind from the X-ray source.

For the low-ionization (zone\,1a) absorber, the material is highly inhomogeneous, as observed from the rapid column changes, and so we estimate the 
mass outflow rate from equation~\ref{eq:mass_outflow_rate}. From the calculations above, $n_{\rm e}\sim10^{8}$\,cm$^{-3}$ and $R\sim10^{18}$\,cm, 
while the volume filling factor is $f_{\rm v}\sim \Delta R/R\sim10^{-4}$ (i.e. the absorber is in the 
form of compact clumps). 
The best estimate of the overall global covering factor comes from the emission associated with this 
zone, which from Table~\ref{tab:xstar} was calculated to be $f_{\rm cov}=0.2$. 
Thus, taking $v_{\rm out}=-0.06c$ and $\mu=1.2$, then the mass outflow rate for zone\,1a is 
$\dot{M}_{\rm out}=1\times 10^{26}$\,g\,s$^{-1}$ or $\sim$1.4 ${\rm M}_{\odot}$\,yr$^{-1}$.

In contrast, as discussed in Section~\ref{sec:two-phase_outflow}, the higher-ionization zone may 
be more homogeneous (and less dense) 
and we thus estimate the mass outflow rate from equation~\ref{eq:mass_outflow_rate_homogeneous}. Taking $N_{\rm H}=10^{22}$\,cm$^{-2}$, 
$R\sim10^{18}$\,cm from above, $v_{\rm out}=-0.06c$ and $f_{\rm cov}=0.45$, as derived from the 
emission (see Table~\ref{tab:xstar}), then 
$\dot{M}_{\rm out}=2\times10^{26}$\,g\,s$^{-1}$ or $\sim$3 ${\rm M}_{\odot}$\,yr$^{-1}$.
Note if the assumption of a homogeneous high-ionization absorber is relaxed, 
as per the variable column density case where the gas is more dense and closer in
(Section~\ref{sec:location_properties_gas}), then $R\sim10^{17}$\,cm, $n_{\rm e}\sim3\times10^{8}$\,cm$^{-3}$ and $\Delta R/R\sim10^{-3}$ for the absorber size-scale. In this case, the mass outflow rate from equation~\ref{eq:mass_outflow_rate} is similar to that obtained for zone\,1a, with $\dot{M}_{\rm out}=6\times10^{25}$\,g\,s$^{-1}$ or $\sim$1 ${\rm M}_{\odot}$\,yr$^{-1}$. 

Although these should only be treated as order of magnitude calculations, even the more conservative estimate above suggests that the combined mass outflow rate for the soft X-ray absorbers 
could still be as high as $\sim$2 ${\rm M}_{\odot}$\,yr$^{-1}$. 
This is similar to what was calculated by \citet{PoundsReeves09} 
for the highly-ionized Fe\,K absorber, of $3.4 {\rm M}_{\odot}$\,yr$^{-1}$, 
assuming a global covering factor of $f_{\rm cov}\sim0.4$, as deduced from the broad ionized 
emission lines present in the archival {\it XMM-Newton} spectra. 
The mass outflow rate either in the soft X-ray band or at Fe\,K 
is also similar to the expected Eddington mass accretion rate 
of PG\,1211+143. For a black hole mass of $10^{8} {\rm M}_{\odot}$ and an efficiency of $\eta=0.1$, 
then $\dot{M}_{\rm Edd}=L_{\rm Edd}/\eta c^{2} \sim2 {\rm M}_{\odot}$\,yr$^{-1}$. 
Furthermore, the outflow momentum rate is $\dot{M} v \sim L_{\rm Edd}/c$, 
which suggests that continuum radiative driving can be important in initially accelerating the gas \citep{KingPounds03}. The higher opacity of the soft X-ray absorber, compared to that at Fe\,K, may then may provide a further boost to the gas acceleration, via an enhanced force multiplier factor.

Note that for a total mass outflow rate of  $\sim$2 ${\rm M}_{\odot}$\,yr$^{-1}$ and $v_{\rm out}=-0.06c$, then the total kinetic power of the soft X-ray outflow is $\dot{E}_{\rm K}=3\times10^{44}$\,erg\,s$^{-1}$.
This is around 3\,per cent of the Eddington luminosity of PG\,1211+143 and, thus, the outflow may be energetically 
significant in terms of galactic feedback \citep{HopkinsElvis10}. In this regard, PG\,1211+143 may be 
similar to other luminous nearby quasars, such as PDS\,456 \citep{Nardini15}, 
IRAS\,F11119+3257 \citep{Tombesi15} and Mrk\,231 \citep{Feruglio15} 
and may give a view of the feedback processes 
that are likely to be common in the early Universe.

An interesting possibility is that the most highly-ionized phase of the outflow, seen in the form of the blueshifted He and H-like Fe\,K absorption lines in the EPIC-pn spectra \citep{Pounds03, PoundsReeves09, Pounds16a}, represents the innermost part of the 
same outflow, launched from a few 100 gravitational radii from the black hole, with the soft X-ray 
wind observed from further out. This may be the case if the wind follows a shallower radial profile than 
$n\propto r^{-2}$, with the ionization then decreasing along the flow \citep{Behar09, Tombesi13}. 
Support for this comes from the similar velocity seen at Fe\,K in the 2014 EPIC-pn data \citep{Pounds16a}, where the dominant component of the Fe\,K outflow has a velocity of $v_{\rm out} = -0.066\pm0.003c$, 
largely consistent with the main components of absorption seen in the soft X-ray band. 

Nonetheless, the existence in 
the soft X-ray band of two distinct ionization phases suggests that we may be viewing a highly inhomogeneous flow. 
Indeed, hydrodynamical disk-wind simulations suggest that the outflows are clumpy and also likely to be highly time-variable \citep{ProgaStoneKallman00, ProgaKallman04, Sim10}, with the wind perhaps fed by relatively short-lived disk-ejection events (Nixon et al. in prep.). In addition, the presence of denser clumps within the wind, as seen here in the form of low-ionization soft X-ray absorbing gas, could play a role in helping to accelerate the fast wind through the extra boost in line-driving 
opacity it subsequently provides \citep{Waters17} and potentially further still if the gas is of relatively high metallicity. 

\section{Conclusions} \label{sec:conclusions}

We have presented an in-depth analysis of the high resolution soft X-ray spectrum of PG\,1211+143, 
taken with {\it XMM-Newton} RGS over a baseline of a month in June 2014. This revealed a rich soft X-ray spectrum, 
with a series absorption lines from He and H-like ions of N, O and Ne, as well as from L-shell Fe, 
systematically blue-shifted by $-0.06c$ and confirming the earlier analysis by \citet{Pounds16b}. Absorption from lower ionization gas is also present, in the form of the Fe M-shell UTA, which also appears blueshifted by $-0.06c$.
 The absorption features can be fitted an absorber with two distinct ionization states, from low and high ionization gas, outflowing at $-0.062\pm0.001c$ and $-0.059\pm0.002c$ respectively. 

These absorption zones were found to be significantly variable, on timescales of just days. The column density was found to vary by more than an order of magnitude, driven by the changes in the Fe M-shell UTA, 
with the overall opacity inversely proportional to the continuum flux. 
The absorber variations 
may be explained in the context of a two phase medium, whereby the variability of the low ionization gas may be explained by compact clumps (of sizescale $\Delta R \sim 10 R_{\rm g}$) passing across 
our line of sight, while the less dense high ionization gas may respond to the continuum in terms 
of its ionization state. The timescales of the variations suggest that the soft X-ray absorbers may be 
located at a typical distance from the black hole of $R\sim10^{17} - 10^{18}$\,cm.
Overall, the observations imply that the outflow in PG\,1211+143 is not a simple homogeneous wind. 
Future calorimeter observations, such as with 
{\it XARM} and {\it Athena}, will be able to further reveal the structure of the wind in PG\,1211+143, 
at high resolution simultaneously from the soft X-rays up to Fe K.

\section{Acknowledgements}

We would like to thank Valentina Braito for providing feedback on the paper.
J.N.\ Reeves acknowledges financial support through NASA grant 
numbers NNX15AF12G, NNX17AC38G and NNX17AD56G. A.\ Lobban acknowledges support from STFC, 
via the consolidated grant ST/K001000/1.
Based on observations obtained with XMM-Newton, an ESA science mission with instruments and contributions directly funded by ESA Member States and NASA.


\begin{thebibliography}

\bibitem[\protect\citeauthoryear{Arnaud}{1996}]{Arnaud96}Arnaud K. A., 1996, in Jacoby G. H., Barnes J., eds, ASP Conf. Ser. Vol. 101, Astronomical Data Analysis Software and Systems V. Astron. Soc. Pac., San Francisco, p. 17

\bibitem[\protect\citeauthoryear{Behar}{2009}]{Behar09}Behar E., 2009, ApJ, 703, 1346

\bibitem[\protect\citeauthoryear{Behar, Sako \& Kahn}{Behar et al.}{2001}]{BeharSakoKahn01}Behar E., Sako M., Kahn S. M., 2001, ApJ, 563, 497

\bibitem[\protect\citeauthoryear{Blustin et al.}{2005}]{Blustin05}Blustin A. J., Page M. J., Fuerst S. V., Branduardi-Raymont G., Ashton C. E., 2005, A\&A, 431, 111

\bibitem[\protect\citeauthoryear{Blustin et al.}{2007}]{Blustin07}Blustin A. J., et al., 2007, A\&A, 466, 107

\bibitem[\protect\citeauthoryear{Boroson \& Green}{1992}]{BorosonGreen92}Boroson T. A., Green R. F., 1992, ApJS, 80, 109

\bibitem[\protect\citeauthoryear{Chartas et al.}{2002}]{Chartas02}Chartas, G., Brandt, W.~N., Gallagher, S.~C., \& Garmire, G.~P.\ 2002, \apj, 579, 169

\bibitem[\protect\citeauthoryear{Costantini et al.}{2007}]{Costantini07}Costantini E., et al., 2007, A\&A, 461, 121

\bibitem[\protect\citeauthoryear{Crenshaw, Kraemer \& George}{Crenshaw et al.}{2003}]{CrenshawKraemerGeorge03}Crenshaw M. D., Kraemer S. B., George I. M., 2003, ARA\&A, 41, 117

\bibitem[Danehkar et al.(2017)]{Danehkar17} Danehkar, A., Nowak, M., Lee, J. C., Kriss, G. A., Young, A., Hardcastle, M. J., Chakrovorty, S., Fang, T., Nielson, J., Rahoui, F., \& Smith, R. K., ApJ, accepted (arXiv:1712.07118)

\bibitem[\protect\citeauthoryear{den Herder et al.}{2001}]{denHerder01}den Herder J. W. et al., 2001, A\&A, 365, 7

\bibitem[\protect\citeauthoryear{Detmers et al.}{2011}]{Detmers11}Detmers R. G., 2011, A\&A, 534, 38

\bibitem[\protect\citeauthoryear{Ferrarese \& Merritt}{2000}]{FerrareseMerritt00}Ferrarese L., Merritt D., 2000, ApJ, 539, 9

\bibitem[\protect\citeauthoryear{Feruglio et al.}{2015}]{Feruglio15}Feruglio C., et al., 2015, A\&A, 583, 99

\bibitem[\protect\citeauthoryear{Gebhardt}{2000}]{Gebhardt00}Gebhardt K., 2000, ApJ, 539, 13

\bibitem[\protect\citeauthoryear{Gehrels et al.}{2004}]{Gehrels04}Gehrels N., et al., 2004, ApJ, 611, 1005

\bibitem[\protect\citeauthoryear{Gofford et al.}{2013}]{Gofford13}Gofford J., Reeves J. N., Tombesi F., Braito V., Turner T. J., Miller L., Cappi M., 2013, MNRAS, 430, 60

\bibitem[\protect\citeauthoryear{Grevesse \& Sauval}{1998}]{GrevesseSauval98}Grevesse N., Sauval A. J., 1998, SSRv, 85, 161

\bibitem[\protect\citeauthoryear{Gupta, Mathur \& Krongold}{Gupta et al.}{2015}]{GuptaMathurKrongold15}Gupta A., Mathur S., Krongold Y., 2015, ApJ, 798, 4

\bibitem[\protect\citeauthoryear{Gupta et al.}{2013}]{Gupta13}Gupta A., Mathur S., Krongold Y., Nicastro F., 2013, ApJ, 772, 66

\bibitem[\protect\citeauthoryear{Halpern}{1984}]{Halpern84}Halpern J. P., 1984, ApJ, 281, 90

\bibitem[\protect\citeauthoryear{Holczer \& Behar}{2012}]{HolczerBehar12}Holczer T., Behar E., 2012, ApJ, 747, 71

\bibitem[\protect\citeauthoryear{Hopkins \& Elvis}{2010}]{HopkinsElvis10}Hopkins P. F., Elvis M., 2010, MNRAS, 401, 7

\bibitem[\protect\citeauthoryear{Jansen et al.}{2001}]{Jansen01}Jansen F. et al., 2001, A\&A, 365, 1

\bibitem[\protect\citeauthoryear{Kaastra et al.}{2000}]{Kaastra00}Kaastra, J.~S., Mewe, R., Liedahl, D.~A., Komossa, S., \& Brinkman, A.~C.\ 2000, \aap, 354, L83 

\bibitem[\protect\citeauthoryear{Kaastra et al.}{2002}]{Kaastra02}Kaastra J. S., Steenbrugge K. C., Raassen A. J. J., van der Meer R. L. J., Brinkman A. C., Liedahl D. A., Behar E., de Rosa A., 2002, A\&A, 386, 427

\bibitem[\protect\citeauthoryear{Kalberla et al.}{2005}]{Kalberla05}Kalberla P. M. W., Burton W. B., Hartmann D., Arnal E. M., Bajaja E., Morras R., P\"{o}ppel W. G. L., 2005, A\&A, 440, 775

\bibitem[\protect\citeauthoryear{Kallman et al.}{1996}]{Kallman96}Kallman T., Liedahl D., Osterheld A., Goldstein W., Kahn S., 1996, ApJ, 465, 994


\bibitem[\protect\citeauthoryear{Kaspi et al.}{2000}]{Kaspi00}Kaspi S., Smith S., Netzer H., Maoz D., Jannuzi B. T., Giveon U., 2000b, ApJ, 533, 631


\bibitem[\protect\citeauthoryear{Kaspi et al.}{2002}]{Kaspi02}Kaspi S., et al., 2002, ApJ, 574, 643

\bibitem[\protect\citeauthoryear{Kaspi et al.}{2004}]{Kaspi04}Kaspi S., Netzer H., Chelouche D., George I. M., Nandra K., Turner T. J., 2004, ApJ, 611, 68

\bibitem[Kaspi \& Behar(2006)]{KaspiBehar06} Kaspi, S., \& Behar, E.\ 2006, \apj, 636, 674 

\bibitem[\protect\citeauthoryear{King}{2003}]{King03}King A. R., 2003, ApJ, 596, L27

\bibitem[\protect\citeauthoryear{King}{2010}]{King10}King A. R., 2010, MNRAS, 402, 1516

\bibitem[\protect\citeauthoryear{King \& Pounds}{2003}]{KingPounds03}King A. R., Pounds K. A., 2003, MNRAS, 345, 657

\bibitem[Kriss et al.(2017)]{Kriss17} Kriss, G. A., Lee, J. C., Danehkar, A., Nowak, M., Fang, T., Hardcastle, M. J., Nielson, J., \& Young, A., ApJ, accepted (arXiv:1712.08850)

\bibitem[\protect\citeauthoryear{Krongold et al.}{2003}]{Krongold03}Krongold Y., Nicastro F., Brickhouse N. S., Elvis M., Liedahl D. A., Mathur S., 2003, ApJ, 597, 832

\bibitem[\protect\citeauthoryear{Lee et al.}{2001}]{Lee01}Lee J. C., Ogle P. M., Canizares C. R., Marshall H. L., Schulz N. S., Morales R., Fabian A. C., Iwasawa K., 2001, ApJ, 554, 13

\bibitem[\protect\citeauthoryear{Lobban et al.}{2016a}]{Lobban16a}Lobban A. P., Pounds K., Vaughan S., Reeves J. N., 2016a, MNRAS, 457, 38

\bibitem[\protect\citeauthoryear{Lobban et al.}{2016b}]{Lobban16b}Lobban A. P., Pounds K., Vaughan S., Reeves J. N., 2016b, ApJ, 831, 201

\bibitem[Lobban et al.(2017)]{Lobban17} Lobban, A. P., Vaughan, S., Pounds, K. A., Reeves, J. N., 2017, MNRAS, submitted

\bibitem[\protect\citeauthoryear{Longinotti et al.}{2010}]{Longinotti10}Longinotti A. L., et al., 2010, A\&A, 510, 92

\bibitem[\protect\citeauthoryear{Longinotti et al.}{2015}]{Longinotti15}Longinotti A. L., Krongold Y., Guainazzi M., Giroletti M., Panessa F., Costantini E., Santos-Lleo M., Rodriguez-Pascual P., 2015, ApJ, 813, 39

\bibitem[\protect\citeauthoryear{Marziani et al.}{1996}]{Marziani96}Marziani P., Sulentic J. W., Dultzin-Hacyan D., Calvani M., Moles M., 1996, ApJS, 104, 37

\bibitem[\protect\citeauthoryear{McKernan, Yaqoob \& Reynolds}{McKernan et al.}{2007}]{McKernanYaqoobReynolds07}McKernan B., Yaqoob T., Reynolds C. S., 2007, MNRAS, 379, 1359


\bibitem[\protect\citeauthoryear{Nahar \& Pradhan}{2003}]{NaharPradhan03}Nahar S. N., Pradhan A. K., 2003, ApJS, 149, 329

\bibitem[\protect\citeauthoryear{Nardini et al.}{2015}]{Nardini15}Nardini E., et al., 2015, Science, 347, 860

\bibitem[Netzer et al.(2003)]{Netzer03} Netzer, H., Kaspi, S., Behar, E., et al.\ 2003, \apj, 599, 933 

\bibitem[\protect\citeauthoryear{Parker et al.}{2017a}]{Parker17} Parker M. L., et al. 2017a, Nature, 543, 83

\bibitem[\protect\citeauthoryear{Parker et al.}{2017b}]{Parker17b} Parker M. L., Reeves J. N., Matzeu G. A., 
Buisson D. J. K., Fabian A. C., 2017, MNRAS, submitted

\bibitem[\protect\citeauthoryear{Peterson et al.}{2004}]{Peterson04}Peterson B., et al., 2004, ApJ, 613, 682

\bibitem[\protect\citeauthoryear{Pinto, Middleton \& Fabian}{Pinto et al.}{2016}]{PintoMiddletonFabian16}Pinto C., Middleton M. J., Fabian A. C., 2016, Nature, 533, 64

\bibitem[\protect\citeauthoryear{Pinto et al.}{2017a}]{Pinto17a}Pinto C., et al., 2017a, MNRAS, submitted; preprint (arXiv:1708.09422)

\bibitem[\protect\citeauthoryear{Pinto et al.}{2017b}]{Pinto17b}Pinto C., et al., 2017b, MNRAS, 468, 2865

\bibitem[\protect\citeauthoryear{Pinto et al.}{2017c}]{Pinto17c}Pinto C., Fabian A., Middleton M., Walton D., 2017c, AN, 338, 234

\bibitem[\protect\citeauthoryear{Pounds}{2014a}]{Pounds14}Pounds K. A., 2014a, MNRAS, 437, 3221

\bibitem[Pounds(2014b)]{Pounds14b} Pounds, K. A.\ 2014b, \ssr, 183, 339

\bibitem[\protect\citeauthoryear{Pounds \& Page}{2006}]{PoundsPage06}Pounds K. A., Page K. L., 2006, MNRAS, 372, 1275

\bibitem[\protect\citeauthoryear{Pounds \& Reeves}{2007}]{PoundsReeves07}Pounds K. A., Reeves J. N., 2007, MNRAS, 374, 823

\bibitem[\protect\citeauthoryear{Pounds \& Reeves}{2009}]{PoundsReeves09}Pounds K. A., Reeves J. N., 2009, MNRAS, 397, 249

\bibitem[\protect\citeauthoryear{Pounds \& Vaughan}{2011}]{PoundsVaughan11}Pounds K. A., Vaughan S., 2011, MNRAS, 413, 1251

\bibitem[Pounds \& King(2013)]{PoundsKing13} Pounds, K.~A., \& King, A.~R.\ 2013, \mnras, 433, 1369 

\bibitem[\protect\citeauthoryear{Pounds et al.}{2001}]{Pounds01}Pounds K., Reeves J., O'Brien P., Page K., Turner M., Nayakshin S., 2001, ApJ, 559, 181

\bibitem[\protect\citeauthoryear{Pounds et al.}{2003}]{Pounds03}Pounds K. A., Reeves J. N., King A. R., Page K. L., O'Brien P. T., Turner M. J. L., 2003, MNRAS, 345, 705

\bibitem[\protect\citeauthoryear{Pounds et al.}{2004}]{Pounds04}Pounds K. A., Reeves J. N., King A. R., Page K. L., 2004, MNRAS, 350, 10

\bibitem[\protect\citeauthoryear{Pounds et al.}{2016a}]{Pounds16a}Pounds K. A., Lobban A., Reeves J. N., Vaughan S., 2016a, MNRAS, 337, 518

\bibitem[\protect\citeauthoryear{Pounds et al.}{2016b}]{Pounds16b}Pounds K. A., Lobban A., Reeves J. N., Vaughan S., Costa M., 2016b, MNRAS, 459, 4389

\bibitem[\protect\citeauthoryear{Proga \& Kallman}{2004}]{ProgaKallman04}Proga D., Kallman T. R., 2004, ApJ, 616, 688

\bibitem[\protect\citeauthoryear{Proga, Stone \& Kallman}{Proga et al.}{2000}]{ProgaStoneKallman00}Proga D., Stone J. M., Kallman T. R., 2000, ApJ, 543, 686

\bibitem[\protect\citeauthoryear{Reeves, O'Brien \& Ward}{Reeves et al.}{2003}]{ReevesOBrienWard03}Reeves J. N., O'Brien P. T., Ward M. J., 2003, ApJ, 593, 65

\bibitem[\protect\citeauthoryear{Reeves et al.}{2013}]{Reeves13}Reeves J. N., Porquet D., Braito V., Gofford J., Nardini E., Turner T. J., Crenshaw D. M., Kraemer S. B., 2013, ApJ, 776, 99

\bibitem[\protect\citeauthoryear{Reeves et al.}{2016}]{Reeves16}Reeves J. N., Braito V., Nardini E., Behar E., O'Brien P. T., Tombesi F., Turner T. J., Costa M. T., 2016, ApJ, 824, 20

\bibitem[\protect\citeauthoryear{Reynolds \& Fabian}{1995}]{ReynoldsFabian95}Reynolds C. S., Fabian A. C., 1995, MNRAS, 273, 1167

\bibitem[\protect\citeauthoryear{Sako et al.}{2001}]{Sako01}Sako M., et al., 2001, A\&A, 365, 168

\bibitem[\protect\citeauthoryear{Sim et al.}{2010}]{Sim10}Sim S. A., Proga D., Miller L., Long K. S., Turner T. J., 2010, MNRAS, 408, 1396

\bibitem[\protect\citeauthoryear{Steenbrugge et al.}{2005}]{Steenbrugge05}Steenbrugge K. C., et al., 2005, A\&A, 434, 569

\bibitem[\protect\citeauthoryear{Tarter, Tucker \& Salpeter}{Tarter et al.}{1969}]{TarterTuckerSalpeter69}Tarter C. C., Tucker W. H., Salpeter E. E., 1969, ApJ, 156, 943

\bibitem[\protect\citeauthoryear{Tombesi et al.}{2010}]{Tombesi10}Tombesi F., Cappi M., Reeves J. N., Palumbo G. G. C., Yaqoob T., Braito V., Dadina M., 2010, A\&A, 521, 57

\bibitem[\protect\citeauthoryear{Tombesi et al.}{2011}]{Tombesi11}Tombesi F., Cappi M., Reeves J. N., Palumbo G. G. C., Braito V., Dadina M., 2011, ApJ, 742, 44

\bibitem[\protect\citeauthoryear{Tombesi et al.}{2013}]{Tombesi13}Tombesi F., Cappi M., Reeves J. N., Nemmen R. S., Braito V., Gaspari M., Reynolds C. S., 2013, MNRAS, 430, 1102

\bibitem[\protect\citeauthoryear{Tombesi et al.}{2015}]{Tombesi15}Tombesi F., Mel\'{e}ndez M., Veilleux S., Reeves J. N., Gonz\'{a}lez-Alfonso E., Reynolds C. S., 2015, Nature, 519, 436

\bibitem[\protect\citeauthoryear{Turner et al.}{2004}]{Turner04}Turner A. K., Fabian A. C., Lee J. C., Vaughan S., 2004, MNRAS, 353, 319

\bibitem[\protect\citeauthoryear{Verner et al.}{1996}]{Verner96}Verner D. A., Ferland G. J., Korista K. T., Yakovlev D. G., 1996, ApJ, 465, 487

\bibitem[\protect\citeauthoryear{Waters et al.}{2017}]{Waters17}Waters T., Proga D., Dannen R., Kallman T. R., 2017, MNRAS, 467, 3160

\bibitem[Willingale et al.(2013)]{Willingale13} Willingale, R., Starling, R. L. C., Beardmore, A. P., Tanvir, N. R., \& O'Brien, P. T. 2013, \mnras, 431, 394

\bibitem[\protect\citeauthoryear{Wilms, Allen \& McCray}{Wilms et al.}{2000}]{WilmsAllenMcCray00}Wilms J., Allen A., McCray R., 2000, ApJ, 542, 914

\end{thebibliography}
\end{document}